\documentclass[10pt]{article}
\usepackage{fullpage}
\usepackage{amsmath}
\usepackage{amssymb}
\usepackage{amsfonts}
\usepackage{comment}
\usepackage{color}
\usepackage{cite}
\usepackage{subcaption}
\usepackage{graphicx}

\def\##1{\underline{#1}}
\def\=#1{\underline{\underline{#1}}}

\def\+
#1{\underline{\bf #1}}
\def\*#1{\underline{\underline{\bf #1}}}

\def\r#1{(\ref{#1})}
\def\l#1{\label{#1}}
\def\c#1{\cite{#1}}

\def\le{\left(}
\def\ri{\right)}
\def\les{\left[}
\def\ris{\right]}
\def\lec{\left\{}
\def\ric{\right\}}

\def\.{\mbox{ \tiny{$^\bullet$} }}

\def\eps{\varepsilon}

\def\epso{\eps_{\scriptscriptstyle 0}}
\def\lambdao{\lambda_{\scriptscriptstyle 0}}
\def\muo{\mu_{\scriptscriptstyle 0}}

\def\ko{k_{\scriptscriptstyle 0}}

\def\ux{\hat{\#u}_x}
\def\uy{\hat{\#u}_y}
\def\uz{\hat{\#u}_z}

\def\calA{{\cal A}}
\def\calB{{\cal B}}

\def\PAmat{\les\=P_\calA\ris}

\def\PBmatz{\les\=P_\calB(z)\ris}

\begin{document}

\begin{center}

\LARGE{ {\bf  Dyakonov--Tamm  surface waves featuring
Dyakonov--Tamm--Voigt  surface waves
}}
\end{center}
\begin{center}
\vspace{10mm} \large

  \vspace{3mm}
 \textbf{Chenzhang Zhou}\\
 {\em NanoMM~---~Nanoengineered Metamaterials Group\\ Department of Engineering Science and Mechanics\\
Pennsylvania State University, University Park, PA 16802--6812, USA} \vspace{3mm}\\
 \textbf{Tom G. Mackay}\footnote{E--mail: T.Mackay@ed.ac.uk.}\\
{\em School of Mathematics and
   Maxwell Institute for Mathematical Sciences\\
University of Edinburgh, Edinburgh EH9 3FD, UK}\\
and\\
 {\em NanoMM~---~Nanoengineered Metamaterials Group\\ Department of Engineering Science and Mechanics\\
Pennsylvania State University, University Park, PA 16802--6812,
USA}
 \vspace{3mm}\\
 \textbf{Akhlesh  Lakhtakia}\\
 {\em NanoMM~---~Nanoengineered Metamaterials Group\\ Department of Engineering Science and Mechanics\\
Pennsylvania State University, University Park, PA 16802--6812, USA}\\
and\\
{\em Danmarks Tekniske Universitet,  Institut for Mekanisk Teknologi, Sektion for Konstruktion og Produktudvikling,
DK-2800 Kongens Lyngby, Danmark}

\normalsize

\end{center}

\begin{center}
\vspace{15mm} {\bf Abstract}
\end{center}

 The propagation of Dyakonov--Tamm (DT) surface waves
guided by the  planar interface of two nondissipative materials $\calA$ and $\calB$ was investigated theoretically and numerically, via the corresponding canonical boundary-value problem. Material $\calA$ is a homogeneous uniaxial dielectric material whose optic axis lies at an angle $\chi$ relative to the interface plane. Material $\calB$ is an isotropic dielectric material that is periodically nonhomogeneous in the direction normal to the interface. The special case was considered in which the propagation matrix for material $\calA$ is non-diagonalizable
 because the  corresponding surface wave~---~named  the Dyakonov--Tamm--Voigt (DTV) surface wave~---~has unusual localization characteristics. The decay of the DTV surface wave  is given by
 the product of a linear function and an exponential function of  distance from the interface in material $\calA$; in contrast,  the fields of conventional DT surface waves  decay only exponentially with distance from the interface.
 Numerical studies revealed that multiple DT surface waves can exist for a fixed propagation direction in the interface plane, depending upon the constitutive parameters of materials $\calA$ and $\calB$.
 When regarded as functions of the angle of propagation in the interface plane, the multiple DT surface-wave  solutions can be organized as  continuous branches.
  A larger number of DT   solution branches exist when the degree of anisotropy of material $\calA$ is greater.  If $\chi = 0^\circ$ then  a solitary  DTV  solution exists for  a  unique propagation direction on each DT branch solution. If $\chi > 0^\circ$, then
 no DTV   solutions exist. As the degree of nonhomogeneity of material $\calB$ decreases, the number of DT  solution branches decreases. 
 For most propagation directions in the interface plane, no  solutions exist in the limiting case wherein the degree of nonhomogeneity approaches zero;
 but
  one
  solution persists provided that the direction of propagation falls within the angular existence domain of the corresponding Dyakonov surface wave.

\section{Introduction}

Electromagnetic surface waves of different types  can be guided by the planar interface of two dissimilar linear materials, depending upon the constitutive characteristics of the two partnering materials \c{Boardman,ESW_book}.
For example, if one partnering material is an isotropic dielectric material and the other is an anisotropic dielectric material, with both materials being homogeneous, then the planar interface can guide the propagation of Dyakonov surface waves \c{MSS,Dyakonov88,Takayama_exp,DSWreview,Walker98}.
A different type of surface wave can propagate if one of the partnering materials is periodically nonhomogeneous in the direction normal to the interface.
For example, if both partnering materials are dielectric materials with one being anisotropic and  one (possibly the same one) being periodically nonhomogeneous,
 then the planar interface can guide the propagation of Dyakonov--Tamm (DT) surface waves \c{LP2007,DT_Reusch_pile}. Both Dyakonov surface waves and DT surface waves can propagate without decay when dissipation is so small that 
 it can be ignored in both partnering materials~---~a characteristics which makes these surface waves attractive for applications involving long-range optical communications \c{Dyakonon1,Dyakonon2}. Unlike Dyakonov surface waves, DT surface waves typically propagate for a wide  range of directions parallel to the interface plane. Also unlike Dyakonov surface waves in the absence of dissipation, multiple DT surface waves 
 with different phase speeds and decay constants
 can propagate in a fixed direction parallel to the interface plane~---~a property which makes them attractive for optical-sensing applications \c{LFjnp}.

All previous works on DT surface waves \c{LP2007,DT_Reusch_pile,LF2,DT_temperature,High_phase_speed_DT}, including experimental observations \c{PML2013,PMLHL2014},
have focused on  the planar interface of a  homogeneous isotropic material and a
periodically nonhomogeneous
 anisotropic material. In contrast, here we consider the planar interface of a
  homogeneous
  anisotropic material and a periodically nonhomogeneous  isotropic material. This case provides a convenient means of studying   Dyakonov--Tamm--Voigt (DTV) surface waves, which have not been described previously.
  
As elaborated upon in the ensuing sections,
a DTV surface wave can exist when
a propagation matrix for the anisotropic partnering material  is non-diagonalizable.
The localization of DTV surface waves 
is fundamentally different from the localization of 
 DT surface waves. Specifically,
as the distance from the planar interface increases in the 
anisotropic partnering material,
the amplitude of a 
 DTV surface wave decays in a combined exponential--linear manner, whereas the amplitudes of
 DT surface waves   decay only in an exponential manner.
  Also, a 
 DTV surface wave propagates  in only a single  direction
 in each quadrant of the interface plane; in contrast,  DT surface waves propagate for a range of directions in each quadrant of the interface plane.

 The fields of the DTV surface wave in the anisotropic partnering material 
have certain characteristics in common with the fields associated with a singular form of 
 planewave propagation called Voigt-wave propagation \c{Voigt,Borzdov,Grundmann}.
 A Voigt wave can exist
   when the planewave propagation matrix is non-diagonalizable \c{Panch,Grech, Ranganath,Gerardin}. Unlike conventional plane waves  \cite{Chen,EAB}, the decay of
  Voigt waves is characterized by
the product of an
exponential function of the propagation distance and a linear function of the
propagation distance.

  In this paper the  theory underpinning the propagation of
DT  and DTV surface waves
is presented for the canonical boundary-value problem  of surface-wave propagation \cite{ESW_book}
 guided by the planar interface of a homogeneous  uniaxial dielectric material and a periodically
 nonhomogeneous isotropic dielectric material. The theory is illustrated by means of representative numerical
calculations, based on realistic values for the constitutive parameters of the partnering materials.

The following notation is adopted:
 The permittivity and permeability of free space are denoted by $\epso$ and $\muo$, respectively. 
 The free-space wavelength is written as $\lambdao = 2 \pi / \ko$ with
 $\ko = \omega \sqrt{\epso \muo}$ being
 the free-space wavenumber 
 and  $\omega$ being  the angular frequency. An $\exp(-i\omega t)$ dependence
 on time $t$ is implicit, with $i = \sqrt{-1}$.
 The
real and imaginary parts of  complex-valued quantities
are delivered by the operators $\mbox{Re} \lec \. \ric$ 
and $\mbox{Im} \lec \. \ric$, respectively. 
Single underlining denotes a 3-vector 
and     $\lec \ux, \uy, \uz \ric$ is
the triad of unit vectors aligned with the Cartesian axes. Dyadics are double underlined \c{Chen}.
Square brackets enclose matrixes and column vectors. The superscript ${}^T$ denotes the transpose. The complex conjugate is denoted by an asterisk.

\section{Theory}

\subsection{Preliminaries}

\begin{figure}[!htb]

\centering
 \includegraphics[width=7.3cm]{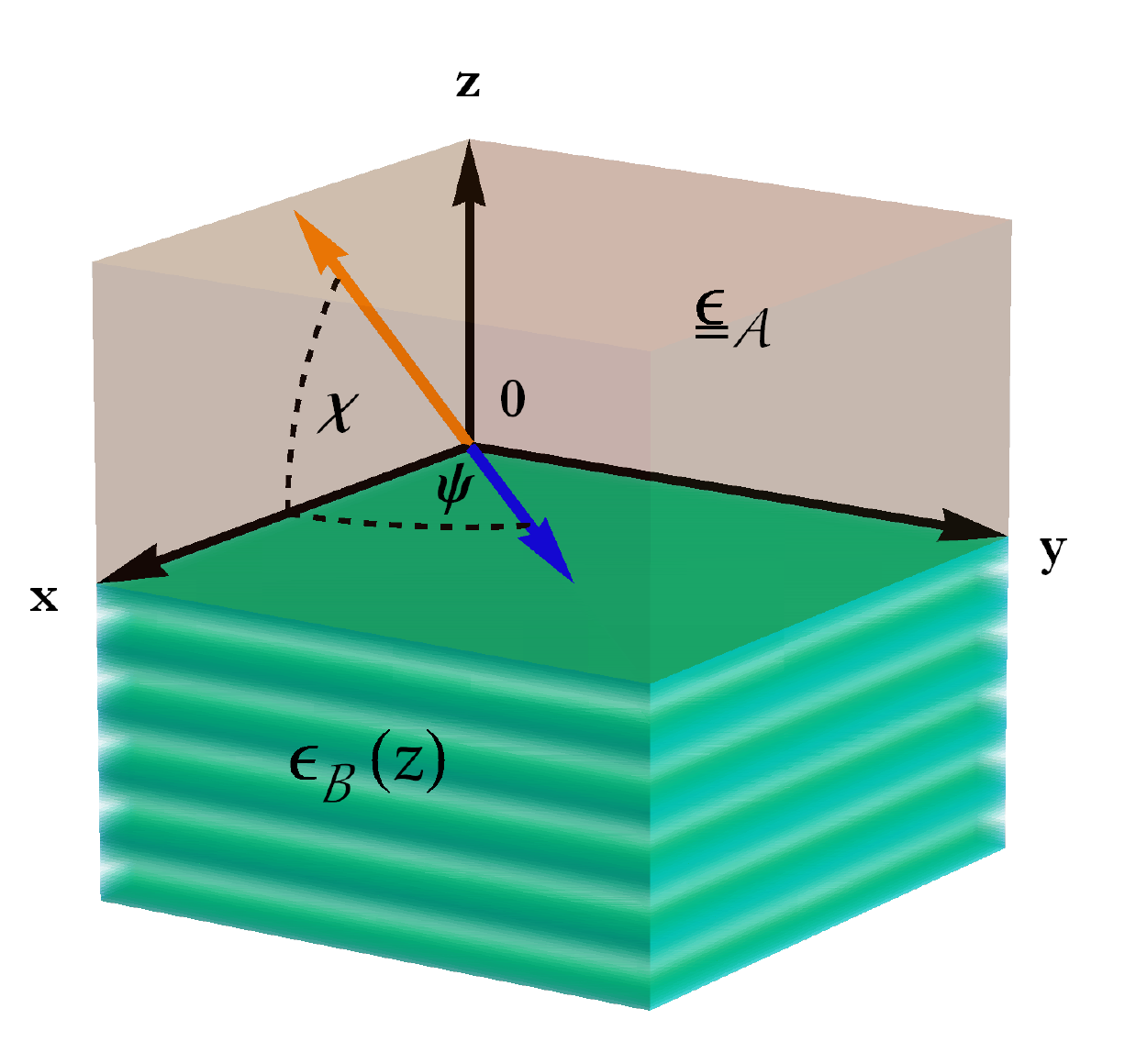} 
  \caption{\label{Fig1} Schematic representation of the canonical boundary-value problem solved. The optic axis of material $\calA$ lies in the $xz$ plane, oriented at the angle $\chi$ relative to $x$-axis. Surface waves propagate parallel to the interface $z=0$, at the angle $\psi$ relative to the $x$-axis. }
\end{figure}

In the canonical boundary-value problem, 
material $\calA$   occupies the half-space $z>0$ and material $\calB$   the half-space $z<0$, as represented schematically in Fig.~\ref{Fig1}.
Whereas material $\calA$ is  anisotropic and homogeneous, material $\calB$ is
isotropic and periodically nonhomogeneous along the $z$ axis. Both materials are dielectric,
and possess neither magnetic nor magnetoelectric properties different from free space \c{ODell,Nye,EAB}.

The relative permittivity dyadic  of material $\calA$ is given as \c{Chen}
\begin{equation}\={\eps}_\calA = \=S_y(\chi) \.
\les {\eps}_\calA^t \ux\, \ux +  {\eps}_\calA^s \le \uy\, \uy + \uz \, \uz \ri \ris  \.
 \=S_y^T(\chi), \end{equation}
wherein the  rotation dyadic
\begin{equation}
\=S_y(\chi) = \uy\,\uy + \le \ux \, \ux  + \uz \, \uz \ri \cos \chi
+\le \uz \, \ux  - \ux \, \uz \ri \sin \chi.
\end{equation}
Thus, the optic axis of material $\calA$ lies wholly in the  $xz$ plane
at an angle $\chi$ with respect to the $x$ axis. Although
the relative permittivity parameters $\eps^s_\calA$ and $\eps^t_\calA$ are generally complex valued,   in the proceeding
numerical studies we have confined ourselves
to $\eps^s_\calA\in \mathbb{R}$ and $\eps^t_\calA \in \mathbb{R}$, as is commonplace in crystal optics \c{BW}.

 The relative permittivity dyadic of material $\calB$
 is specified as $\=\eps_\calB(z)=\eps_\calB(z)\=I$, where
\begin{equation} \l{eB}
\eps_\calB(z) = 
\left[\frac{n_1+n_2}{2}+ \gamma \frac{n_1-n_2}{2}\sin\left(\frac{\pi z}{\Omega}\right)
\right]^2
\,,
\end{equation}
and $\=I=\ux\, \ux +
  \uy\, \uy + \uz \, \uz$ is the 3$\times$3 identity dyadic  \cite{Chen}. 
 In Eq.~\r{eB}, the parameter $\Omega > 0$ 
 is the half-period  of the periodic variation in  dielectric properties along the negative $z$ axis, while $\gamma > 0$ is a
  scaling parameter for the
 amplitude of this periodic variation.
  The parameter $\gamma$ can be considered to be the degree of nonhomogeneity when $\Omega$ is finite.
  We take the refractive indexes $n_1$ and $n_2$ to be real and positive,
  as is commonplace in the literature on rugate filters \cite{Bovard,Bau}.

The electromagnetic field phasors for surface-wave propagation 
are expressed everywhere as
\cite{ESW_book} 
\begin{equation} \label{planewave}
\left.\begin{array}{l}
 \#E (\#r)=  \les e_x(z)\ux + e_y(z)\uy+e_z(z)\uz \ris \, 
 \exp\les i q \le x \cos \psi + y \sin \psi \ri \ris \\[4pt]
 \#H  (\#r)=  \les h_x(z)\ux + h_y(z)\uy+h_z(z)\uz \ris \,
 \exp\les i q \le x \cos \psi + y \sin \psi \ri \ris 
 \end{array}\right\}\,,  \:\:\: -\infty < z < + \infty,
\end{equation}
with ${q}$ being the surface  wavenumber. The angle $\psi\in\left[0,2\pi\right)$
specifies
 the direction of propagation in the $xy$ plane, relative to  the $x$ axis.
 The phasor representations~\r{planewave}, when combined with
the source-free Faraday and Amp\'ere--Maxwell equations,  deliver 
 the 4$\times$4 matrix ordinary differential
equations \c{Billard,Berreman}
\begin{equation}
\label{MODE_A}
\frac{d}{dz}\les\#f  (z)\ris= \left\{
\begin{array}{l}
i \PAmat\.\les\#f  (z)\ris\,,  \qquad   z>0 \vspace{8pt} \\
i \PBmatz\.\les\#f  (z)\ris\,,  \qquad   z<0
\end{array} , \right.
\end{equation}
wherein
the  column 4-vector
\begin{equation}
\les\#f (z)\ris= 
\les
\begin{array}{c}%
e_ x(z), \quad
e_y(z),\quad
h_x(z),\quad
h_y(z)
\end{array}
\ris^T,
\label{f-def}
\end{equation}
and  the 4$\times$4 propagation matrixes  $\PAmat$ and $\PBmatz$   are determined by  $\=\eps_\calA$ and $\eps_\calB(z)$, respectively.
The $x$-directed and $y$-directed components of the phasors
are algebraically connected to their   $z$-directed components  \c{EAB,FL2010}.

\subsection{Half-space $z>0$}

The  4$\times$4 propagation matrix  $\les\=P_\calA\ris $ is given as \cite{ZML_JOSAB}
\begin{equation}
\les\=P_\calA\ris=  \les   
\begin{array}{cccc}
\displaystyle{\frac{\beta}{ \Gamma }}&0& 
\displaystyle{\frac{\tau}{ \omega \epso \Gamma }} & 
\displaystyle{ \frac{k_o^2 \Gamma- \nu_c}{\omega \epso \Gamma}  } \vspace{6pt} \\
\displaystyle{\frac{\beta \tan \psi }{ \Gamma}}&0& 
 \displaystyle{\frac{\nu_s -k_o^2 \Gamma }{\omega \epso \Gamma}}&
\displaystyle{- \frac{\tau}{ \omega \epso \Gamma}}  \vspace{6pt} 
\\
 \displaystyle{-\frac{\tau}{\omega \muo}} & 
\displaystyle{\frac{\nu_c -\ko^2 \eps^s_\calA}{\omega \muo}}  &0&0  \vspace{6pt}  \\
\displaystyle{\frac{\ko^2 \eps^s_\calA \eps^t_\calA- \Gamma \nu_s }{\omega \muo \Gamma }}  &
\displaystyle{\frac{\tau}{\omega \muo}}&
- \displaystyle{\frac{\beta \tan\psi}{ \Gamma }}&
\displaystyle{\frac{\beta }{ \Gamma }}
\end{array}
\ris, 
\end{equation}
wherein the generally complex-valued parameters 
\begin{equation}
\left.
\begin{array}{l}
\nu_c = q^2  \cos^2 \psi \vspace{4pt} \\
\nu_s = q^2  \sin^2 \psi \vspace{4pt} \\
\beta=  q \le \eps^s_\calA- \eps^t_\calA  \ri \sin \chi \cos \chi \cos \psi  \vspace{4pt} \\
\Gamma=  \eps^s_\calA \cos^2 \chi + \eps^t_\calA \sin^2 \chi  \vspace{4pt}\\
\tau = q^2 \cos \psi \sin \psi
\end{array}
\right\}.
\end{equation}
The    $z$-directed  components of the field phasors are
\begin{equation} \l{z_comp}
\left.
\begin{array}{l}
e_z(z) = \displaystyle{\frac{1}{\Gamma} \left\{\frac{q \les h_ x(z) \sin \psi - h_ y(z)  \cos \psi  \ris}{\omega \epso } \right.} 
+ \displaystyle{ e_x(z) \le \eps^s_\calA - \eps^t_\calA  \ri \sin \chi \cos \chi \Bigg\}}  \vspace{8pt} \\
h_ z(z) = \displaystyle{\frac{q \les e_ y(z) \cos \psi - e_ x(z)  \sin \psi  \ris}{\omega \muo }}
\end{array}
\right\}\,,\qquad z > 0\,.
\end{equation}

\subsubsection{Dyakonov--Tamm surface wave}

Before dealing with DTV surface waves, it is necessary  to first consider
DT surface waves for which  $\les\=P_\calA\ris$ has four eigenvalues, 
each with algebraic multiplicity $1$ and geometric multiplicity $1$. 
These eigenvalues are \cite{ZML_JOSAB}
\begin{equation} \l{a_decay_const}
\left.
\begin{array}{l}
\alpha_{\calA a} = i \sqrt{ q^2 - \ko^2 \eps_\calA^s} \vspace{8pt}\\
\alpha_{\calA b} = - i \sqrt{ q^2 - \ko^2 \eps_\calA^s} \vspace{8pt}\\
\alpha_{\calA c} = \displaystyle{\frac{ \beta + i  \sqrt{ \eps^s_\calA \les  \nu_s \cos^2 \chi \le  \eps^s_\calA -  \eps^t_\calA \ri + q^2 \eps^t_\calA \ris -  \Gamma \eps^s_\calA \eps^t_\calA k_o^2 } }{\Gamma} }
 \vspace{8pt}\\
\alpha_{\calA d} = \displaystyle{\frac{  \beta - i  \sqrt{ \eps^s_\calA \les  \nu_s \cos^2 \chi \le  \eps^s_\calA -  \eps^t_\calA \ri + q^2 \eps^t_\calA \ris -  \Gamma \eps^s_\calA \eps^t_\calA k_o^2 }}{\Gamma}}
\end{array}
\right\}\,.
\end{equation}
Eigenvalues which have negative  imaginary parts are irrelevant for  surface-wave propagation \cite{ESW_book}.
Either
 $ \alpha_{\calA a} $ 
   $ \alpha_{\calA b} $ can have a positive imaginary part, but both cannot.
Let us also assume that  
 only one of 
 $ \alpha_{\calA c} $ 
 and $ \alpha_{\calA d} $ can have a positive  imaginary part. Therefore the two eigenvalues that are chosen \cite{ZML_JOSAB} for  surface-wave analysis are 
 \begin{equation} \l{alphaA1}
 \begin{array}{l}
 \alpha_{\calA 1} = \left\{ 
 \begin{array}{lr}
 \alpha_{\calA a} & \quad \mbox{if} \quad \mbox{Im} \lec 
 \alpha_{\calA a} \ric > 0 \vspace{0pt}\\
 \alpha_{\calA b} & \quad  \mbox{otherwise}
 \end{array}
 \right. 
 \end{array}
 \end{equation}
 and
  \begin{equation} \l{alphaA2}
 \begin{array}{l}
 \alpha_{\calA 2} = \left\{ 
 \begin{array}{lr}
 \alpha_{\calA c} & \quad \mbox{if} \quad \mbox{Im} \lec 
 \alpha_{\calA c} \ric > 0 \vspace{0pt}\\
 \alpha_{\calA d} & \quad  \mbox{otherwise}
 \end{array}
 \right. 
 \end{array}.
 \end{equation}
 
 Explicit expressions for the corresponding eigenvectors $\les\#v_{\calA 1}\ris$
 and $\les\#v_{\calA 2}\ris$  can be derived by solving the equations
\begin{equation}
\le \les\=P_\calA\ris - \alpha_{\calA  1} \les\=I\ris \ri \. \les\#v_{\calA 1} \ris=\les \#0\ris
\end{equation}
and
\begin{equation}
\le \les\=P_\calA\ris - \alpha_{\calA  2} \les\=I\ris \ri \. \les\#v_{\calA 2} \ris=\les \#0\ris,
\end{equation}
where $\les\=I\ris$ is the 4$\times$4 identity matrix and $\les\#0\ris$ is the null column 4-vector, but the
expressions  are too cumbersome for reproduction here. More importantly,  the general solution of Eq.~\r{MODE_A}${}_1$  representing DT surface waves that decay as $z \to +\infty$ is given as 
 \begin{equation} \l{D_gen_sol}
\les\#f  (z)\ris =  C_{\calA 1}  \les\#v_{\calA 1}\ris  \exp \le i \alpha_{\calA 1 } z \ri
+ C_{\calA 2}   \les\#v_{\calA 2} \ris  \exp \le i \alpha_{\calA 2} z \ri
\,,\quad z > 0\,.
\end{equation}
The complex-valued constants $C_{\calA 1}$ and $C_{\calA 2}$ herein are fixed 
by applying  boundary conditions at $z=0$. These boundary conditions involve
 \begin{equation} \l{bcD+}
\les\#f  (0^+)\ris =  C_{\calA 1}  \les\#v_{\calA 1}\ris   
+ C_{\calA 2}   \les\#v_{\calA 2} \ris  
\,.
\end{equation}

\subsubsection{Dyakonov--Tamm--Voigt surface wave}

We must have $\alpha_{\calA 1} = \alpha_{\calA 2} = \alpha_\calA$  for DTV surface-wave propagation. Thus,
 $\les\=P_\calA\ris$ has only two eigenvalues, 
each with algebraic multiplicity $2$ and geometric multiplicity $1$. 
There are  four possible values of $q$ that result in $\alpha_{\calA 1} = \alpha_{\calA 2}$, namely \cite{ZML_JOSAB}
\begin{equation}
q = 
\left\{
\begin{array}{l}
\displaystyle{
\frac{\ko \sqrt{\eps^s_\calA} \cos \chi \le \cos \psi \pm i \sin \chi \sin \psi \ri}{1 - \cos^2 \chi \sin^2 \psi}} \vspace{8pt} \\
\displaystyle{
- \frac{\ko \sqrt{\eps^s_\calA} \cos \chi \le \cos \psi \pm i \sin \chi \sin \psi \ri}{1 - \cos^2 \chi \sin^2 \psi}}
\end{array}
\right. ,
\end{equation}
 with 
the  correct value of $q$ for DTV surface-wave propagation being the one that yields 
$\mbox{Im} \lec \alpha_{\calA} \ric > 0 $ \cite{ESW_book,ZML_JOSAB}.
Although explicit expressions for a corresponding eigenvector $\les\#v_\calA\ris$ satisfying
\begin{equation}
\le \les\=P_\calA\ris - \alpha_{\calA } \les\=I\ris \ri \. \les\#v_{\calA } \ris=\les \#0\ris,
\end{equation}
and a corresponding generalized eigenvector $\les\#w_\calA\ris$ satisfying
 \cite{Boyce}
\begin{equation}
\le \les\=P_\calA\ris - \alpha_{\calA } \les\=I\ris \ri \. \les\#w_{\calA }\ris = \les\#v_{\calA }\ris,
\end{equation}
were derived, the expressions are too cumbersome to be reproduced here. 

Thus, the general solution of Eq.~\r{MODE_A}${}_1$  representing DTV surface waves that decay as $z \to +\infty$
can be stated as
\begin{equation} \l{DV_gen_sol}
\les\#f  (z)\ris = \Big( C_{\calA 1}  \les\#v_{\calA }\ris  + C_{\calA 2} \lec i  z \, \les\#v_{\calA} \ris  + \les\#w_{\calA}\ris \ric  \Big) \exp \le i \alpha_{\calA } z \ri\,,\quad z > 0\,.
\end{equation}
The complex-valued constants $C_{\calA 1}$ and $C_{\calA 2}$ herein are fixed 
by applying  boundary conditions at $z=0$. These boundary conditions involve
 \begin{equation} \l{bcDV+}
\les\#f  (0^+)\ris =  C_{\calA 1}  \les\#v_{\calA }\ris   
+ C_{\calA 2}   \les\#w_{\calA } \ris  
\,.
\end{equation}

\subsection{Half-space $z<0$}

The  4$\times$4 propagation matrix $\PBmatz$   is given as \cite{ESW_book,FL2010}
\begin{eqnarray}
\PBmatz
\les   
\begin{array}{cccc}
0&0& \displaystyle{ \frac{\tau}{\omega \epso \eps_\calB(z)}} & 
\displaystyle{\frac{\ko^2 \eps_\calB(z)- \nu_c }{\omega \epso \eps_\calB(z)} } \vspace{8pt} \\
0&0& \displaystyle{\frac{ \nu_s -\ko^2 \eps_\calB (z)}{\omega \epso \eps_\calB(z)} }&
\displaystyle{ -\frac{\tau}{\omega \epso \eps_\calB(z)}} \vspace{8pt}
\\
\displaystyle{ -\frac{\tau}{\omega \muo}} & 
\displaystyle{\frac{\nu_c -\ko^2 \eps_\calB(z)}{\omega \muo} } &0&0\vspace{8pt} \\
\displaystyle{\frac{\ko^2 \eps_\calB(z)- \nu_s}{\omega \muo} } &
\displaystyle{ \frac{\tau}{\omega \muo}}&0&0
\end{array}\ris. 
\end{eqnarray}
The $z$-directed components of the  phasors are given by 
\begin{equation}
\left.
\begin{array}{l}
e_ z(z) = \displaystyle{\frac{q \les h_ x(z) \sin \psi - h_ y(z)  \cos \psi  \ris}{\omega \epso 
\eps_\calB (z) }} \vspace{8pt} \\
h_ z(z) = \displaystyle{\frac{q \les e_ y(z) \cos \psi - e_ x(z)  \sin \psi  \ris}{\omega \muo }}
\end{array}
\right\}\,,\qquad z < 0\,.
\end{equation}

Equation~\r{MODE_A}${}_2$ has to be solved numerically, even though
the form of its solution known by virtue of the Floquet--Lyapunov theorem \cite{Hochstadt,YS75}.
 The optical response of one period of material $\calB$ for specific
values of $q$ and $\psi$
is characterized by the matrix $[\=Q_\calB]$ that appears in the relation
\begin{equation}\label{Q_B}
[\#f(z)]=[\=Q_\calB]\.[\#f(z-2\Omega)]\,, \qquad z < 0\,.
\end{equation}
A matrix $[\=A_\calB]$ is defined through the following relation:
\begin{equation}
[\=Q_\calB] = \exp\left\{i2\Omega[\=A_\calB]\right\}\,.
\end{equation}
Both $[\=Q_\calB]$ and $[\=A_\calB]$ share the same (linearly independent) eigenvectors, and their eigenvalues
are also related. Let $[\#v_{\calB n}]$, $n\in\left[1,4\right]$, be the eigenvector corresponding
to the  $n$th eigenvalue $\sigma_{\calB n}$ of $[\=Q_\calB]$; then, the corresponding eigenvalue
$\alpha_{\calB n}$ of $[\=A_\calB]$
is given by
\begin{equation}
\alpha_{\calB n} = -i\frac{\ln \sigma_{\calB n}}{2\Omega}\,,\qquad n\in\left[1,4\right]\,.
\end{equation}
After labeling the eigenvalues of  $[\=A_\calB]$ such that
$\mbox{Im}\lec {\alpha_{\calB 3}} \ric<0$ and $\mbox{Im}\lec {\alpha_{\calB 4}} \ric<0$,
we set  
\begin{equation}
[\#f(0^-)]= C_{\calB 3}  \les\#v_{\calB 3}\ris   
+ C_{\calB 4}   \les\#v_{\calB 4} \ris 
\end{equation}
for surface-wave propagation,
where the complex-valued constants $C_{\calB 3}$ and $C_{\calB 4}$ are fixed 
by applying  boundary conditions at $z=0$.
The other two eigenvalues of  $[\=A_\calB]$ pertain to waves that amplify as $z\to-\infty$
and cannot therefore contribute to the surface wave.

The piecewise-uniform-approximation method is used to calculate $[\=Q_\calB]$, and thereby $\les \#f(z) \ris$ for all $z< 0$, as follows \cite{ESW_book}. The $z<0$ half-space is partitioned in to   slices of equal thickness, with each cut occurring at the plane $z=z_n$ where
\begin{equation}
z_n = \frac{2 \Omega n }{N}
\end{equation}
for all integers $n \in \le -\infty, -1 \ris$, the integer $N>0$ being the number of slices per period  along the negative $z$ axis. 
The matrixes
\begin{equation}
\les\=W_\calB\ris^{(n)} =
\exp\left\{ i \le z_n - z_{n+1} \ri  \les\=P_\calB(\frac{z_{n+1} + z_n}{2})\ris\right\},
\qquad n \in \le -\infty, -1 \ris,
\end{equation}
are introduced.
 As propagation from the plane $z = z_{n+1}$ to the plane $z=z_n$ is characterized approximately by the matrix $\les\=W_\calB\ris^{(n)}$,
we get
\begin{equation}
 [\=Q_\calB] \cong \les\=W_\calB\ris^{(N)} \. \les\=W_\calB\ris^{(N-1)} \. \cdots \. \les\=W_\calB\ris^{(2)} \.\les\=W_\calB\ris^{(1)}.
\end{equation}
The integer $N$ should be sufficiently large so that the piecewise-uniform approximation   captures well the continuous variation of $\les\=P_\calB(z) \ris$. 
The piecewise-uniform approximation
to $[\#f(z)]$ for arbitrary $z<0$ is accordingly given by
\begin{equation}
[\#f(z)] \cong
\left\{
\begin{array}{l}
\displaystyle{\exp\left\{ i  z    \les\=P_\calB(\frac{z_{-1} }{2})\ris\right\}\. [\#f(0^-)]},\qquad z \in \les z_{-1},0 \ri,  \vspace{8pt} \\
\displaystyle{
\exp\left\{ i \le z- z_{n} \ri  \les\=P_\calB(\frac{z_{n-1} + z_n}{2})\ris\right\} \. \les\=W_\calB\ris^{(n)} 
\vspace{8pt}\. \les\=W_\calB\ris^{(n+1)} \. \cdots
} \\
\displaystyle{
  \. \les\=W_\calB\ris^{(-2)} \.\les\=W_\calB\ris^{(-1)} \. [\#f(0^-)]},\qquad z \in \les z_{n-1},z_{n} \ris, \qquad n \in \le - \infty, -1 \ris.
\end{array}
\right. 
\end{equation}

\subsection{Application of boundary conditions}

 The continuity of the tangential  components of the electric and magnetic field
 phasors across the interface plane $z=0$ imposes
 four conditions that are represented compactly as
 \begin{equation}
 \label{2.23-AL}
 \les\#f(0^+)\ris=  \les\#f(0^-)\ris
 \,.
 \end{equation}
Accordingly,
\begin{equation}
\les \=Y \ris \. \les \:
 C_{\calA 1}, \quad
  C_{\calA 2}, \quad
   C_{\calB 3}, \quad
    C_{\calB 4} \:
 \ris^T =  \les\#0\ris\,,
\end{equation}
wherein the 4$\times$4 characteristic matrix $\les \=Y \ris$ must be singular for  surface-wave propagation \c{ESW_book}.
The dispersion equation 
\begin{equation}
\l{dispersion_eq}
\left\vert \les \=Y \ris\right\vert = 0,
 \end{equation}  can be numerically solved for $q$ for a fixed value
 of $\psi$, by the Newton--Raphson method \c{N-R} for example .

 \section{Numerical results and discussion}
 
 The solutions of the dispersion equation \r{dispersion_eq} were explored numerically for $\lambdao=633$~nm.
Constitutive parameters
 corresponding to a realistic rugate filter \c{Bovard,Bau}
  were chosen for material $\cal B$: $n_1 =  2.32$, $n_2=1.45$, and $\Omega=5 \lambdao$. Whereas
  $\eps^t_\calA=4$ was fixed,  $\eps^s_\calA$ was kept variable in order to ensure the excitation
  of DTV surface waves. Parenthetically, regimes involving larger values of the half-period prove to be inaccessible 
due to a loss of numerical stability \c{FL2010}.
 
\begin{figure}[!htb]

\centering
\begin{subfigure}{0.49\textwidth}
 \includegraphics[width=7.3cm]{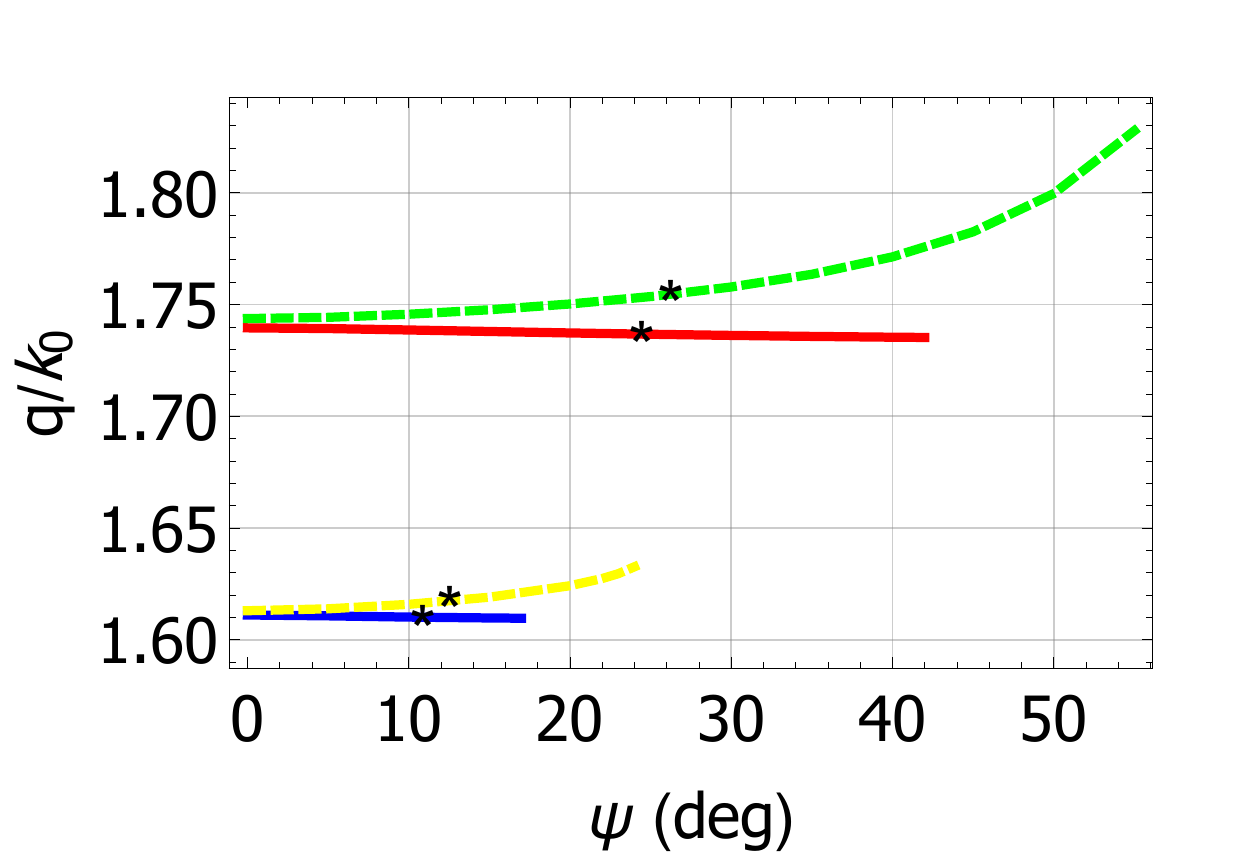} 
 \caption{}
 \end{subfigure} \hfill
 \begin{subfigure}{0.49\textwidth}
  \includegraphics[width=7.3cm]{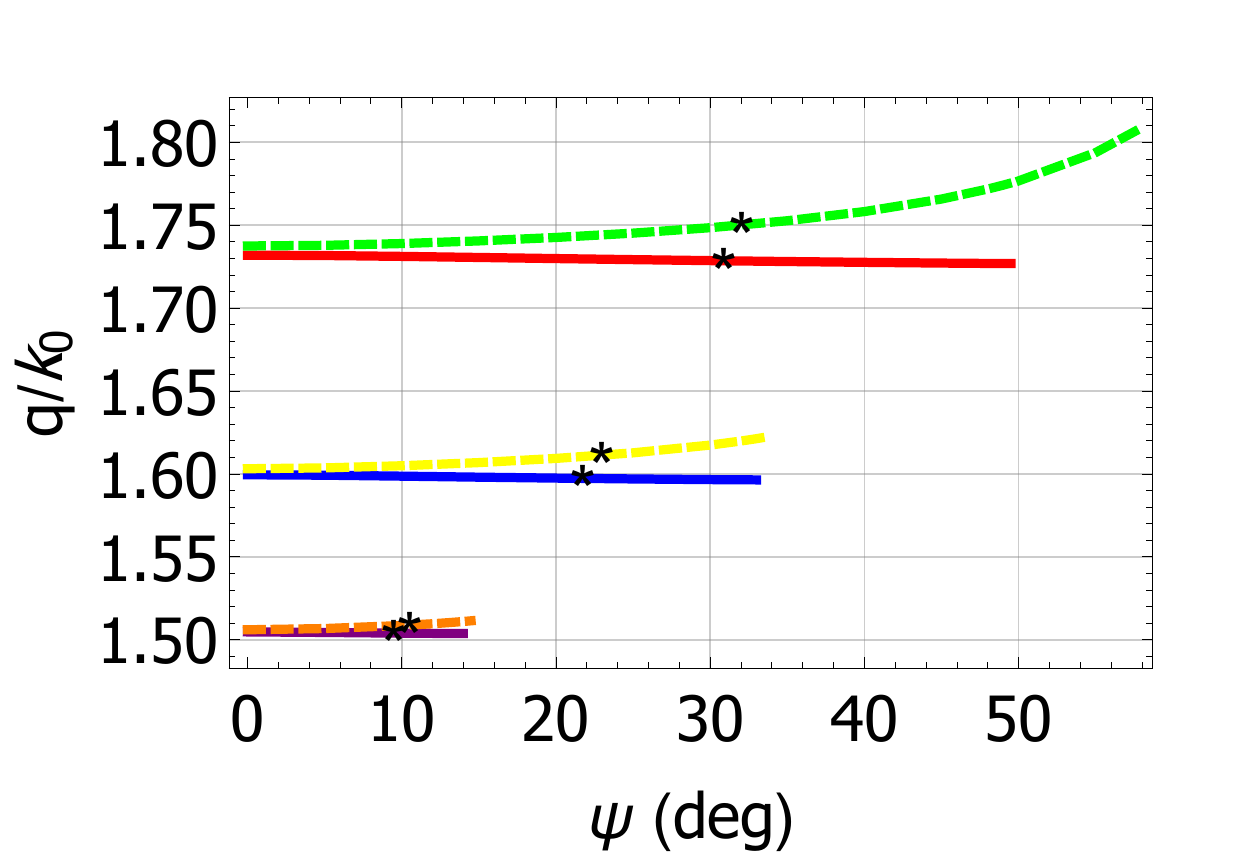}
 \caption{}
  \end{subfigure}\\
  \begin{subfigure}{0.49\textwidth}
     \includegraphics[width=7.3cm]{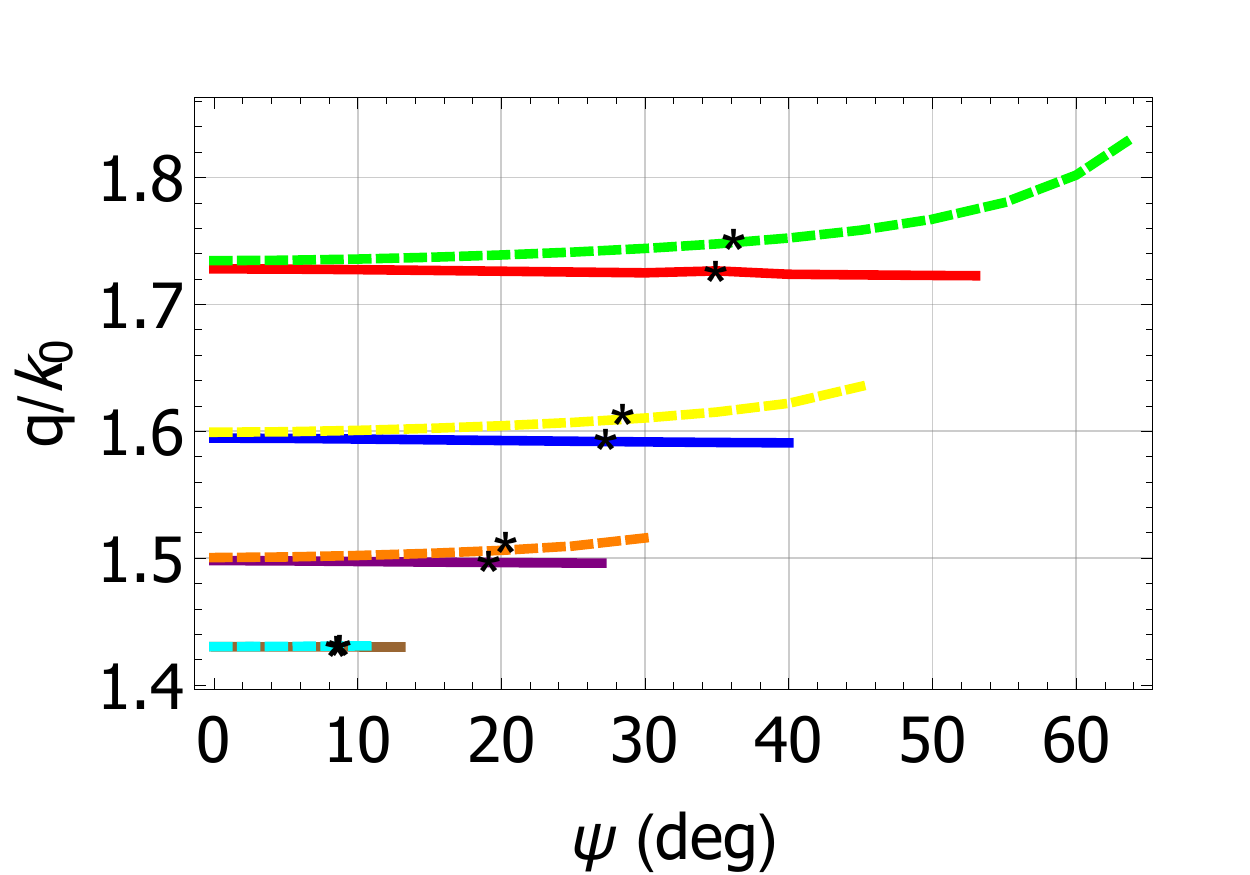} 
     \caption{}
     \end{subfigure}
  \caption{\label{Fig2}   
Plots of  $q/\ko $
versus 
$\psi $ 
for  $\chi=0^\circ$ and $\gamma=1$,  when (a) $\eps^s_\calA=2.5$,  (b) $\eps^s_\calA=2.2,$ and   (c) $\eps^s_\calA=2$. 
The  curves represent DT surface-wave solutions: there are 4 branches for (a), 
6  for (b), and 8   for (c). On each   branch,
 the corresponding  DTV surface-wave solution is identified by a star.
    }
\end{figure}

 In Figs.~\ref{Fig2}(a-c) plots are provided of
   $q/\ko $ versus 
$\psi $, as obtained from Eq.~\r{dispersion_eq}.
 For these calculations
  $\chi=0^\circ$ and $\gamma=1$,  with (a) $\eps^s_\calA=2.5$,  (b) $\eps^s_\calA=2.2,$ and   (c) $\eps^s_\calA=2$. 
  Representing DT surface waves, the solutions organized as branches:  there are 4 branches for $\eps^s_\calA=2.5$, 
6   for $\eps^s_\calA=2.2$, and 8  for $\eps^s_\calA=2$.
 Each branch   exists for a continuous range of $q$, say $  q_{\text{min}} < q <   q_{\text{max}}$ and a continuous range of $\psi$, say $\psi_{\text{min}} < \psi < \psi_{\text{max}}$.
For every branch,  $\psi_{\text{min}} = 0^\circ$, while $\psi_{\text{max}} \in \le 11^\circ, 64^\circ \ri$ depending on the value of $\eps^s_\calA$.
The   solution branches that arise at higher values of $ q_\text{min}$
exist for wider ranges of values of $\psi$. The value of
$q$ on each branch   increases slowly as $\psi$ increases towards $\psi_{\text{max}}$.

 On every DT branch in Figs.~\ref{Fig2}(a-c), for a unique value  of $\psi$ and a unique value of $q$,
 there exists a 
   DTV surface-wave solution~---~which is represented by a star.
   These DTV solutions do not arise at $\psi \in\lec \psi_{\text{min}},\psi_{\text{max}}\ric$ nor at  $q  \in\lec   q _{\text{min}},  q _{\text{max}}\ric$; instead, they arise at mid-range values of $\psi$ and $q$.

\begin{figure}[!htb]

\centering
 \includegraphics[width=7.3cm]{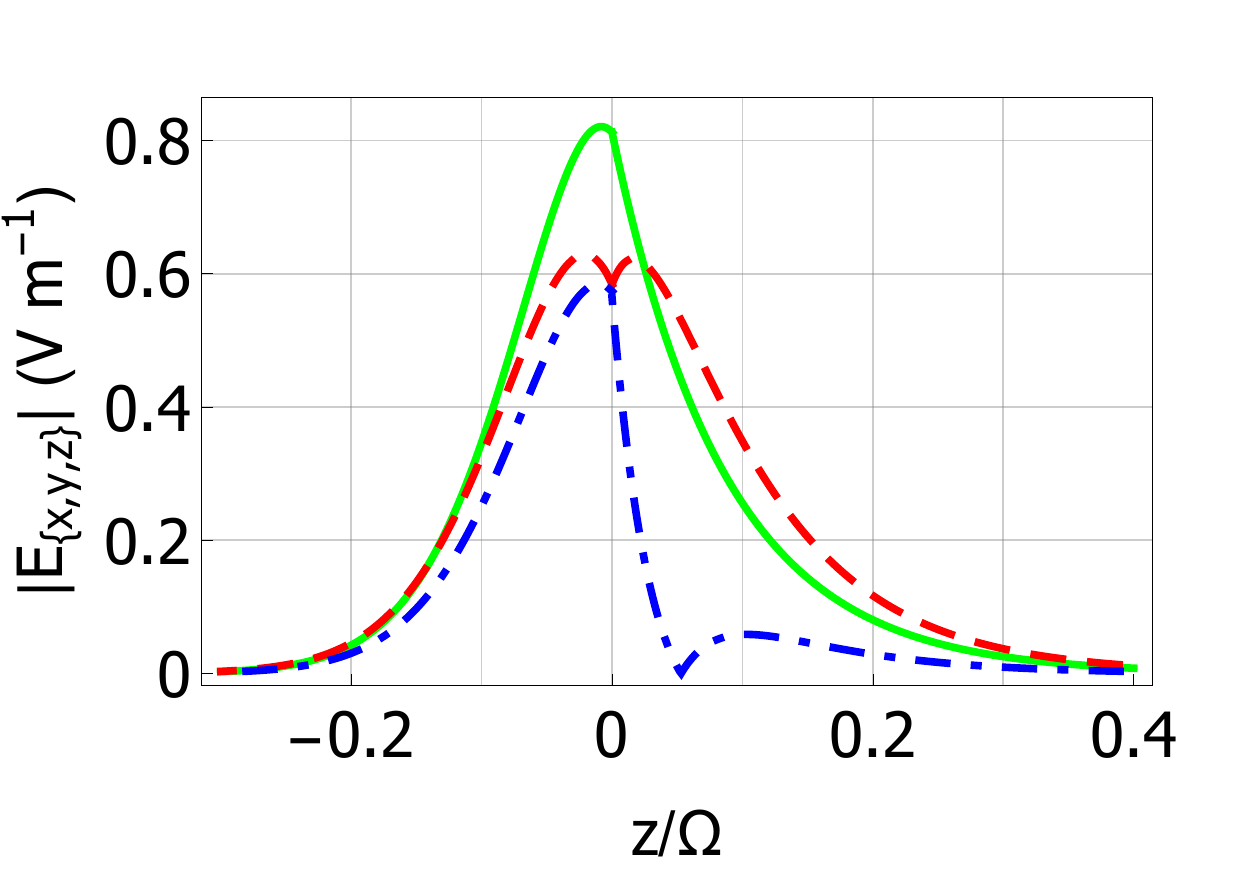}  \includegraphics[width=7.3cm]{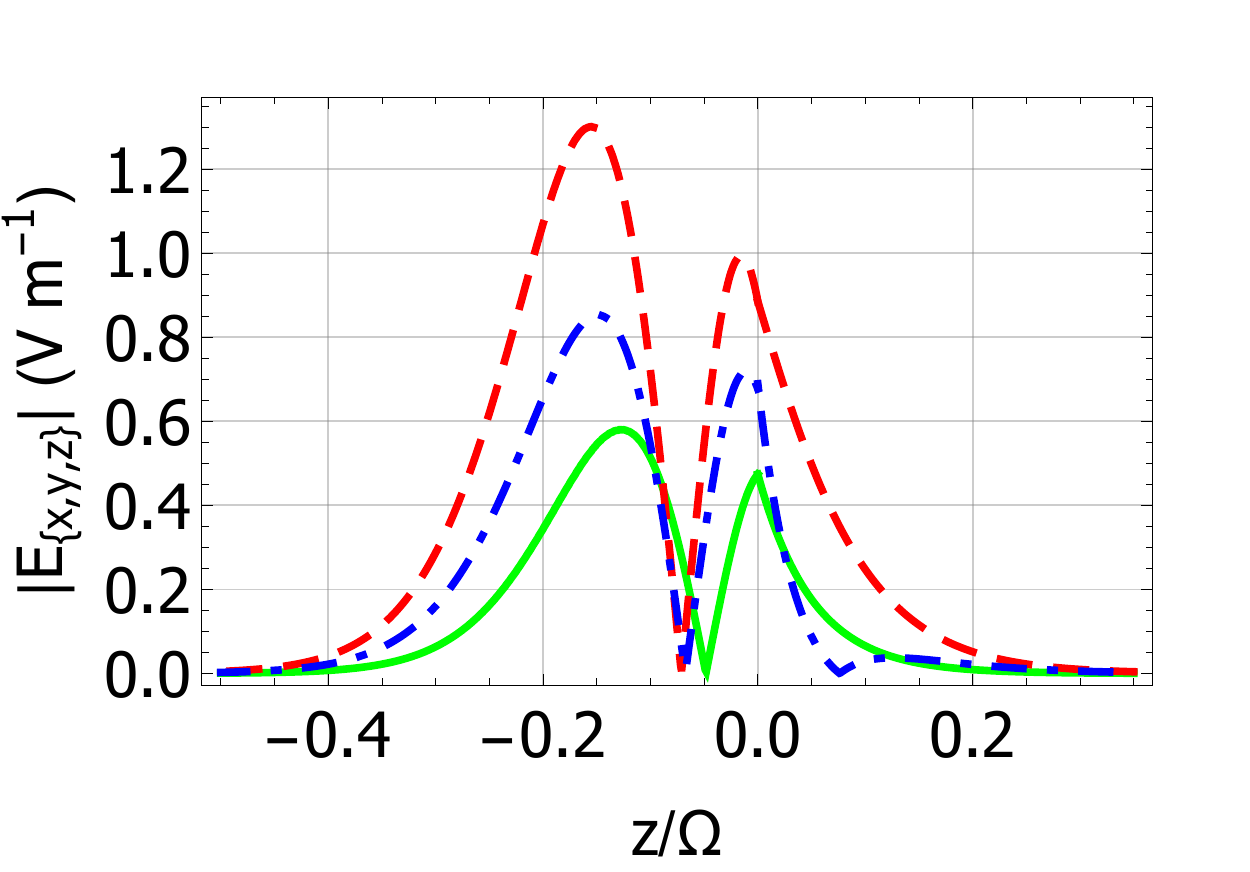}  \\
 \includegraphics[width=7.3cm]{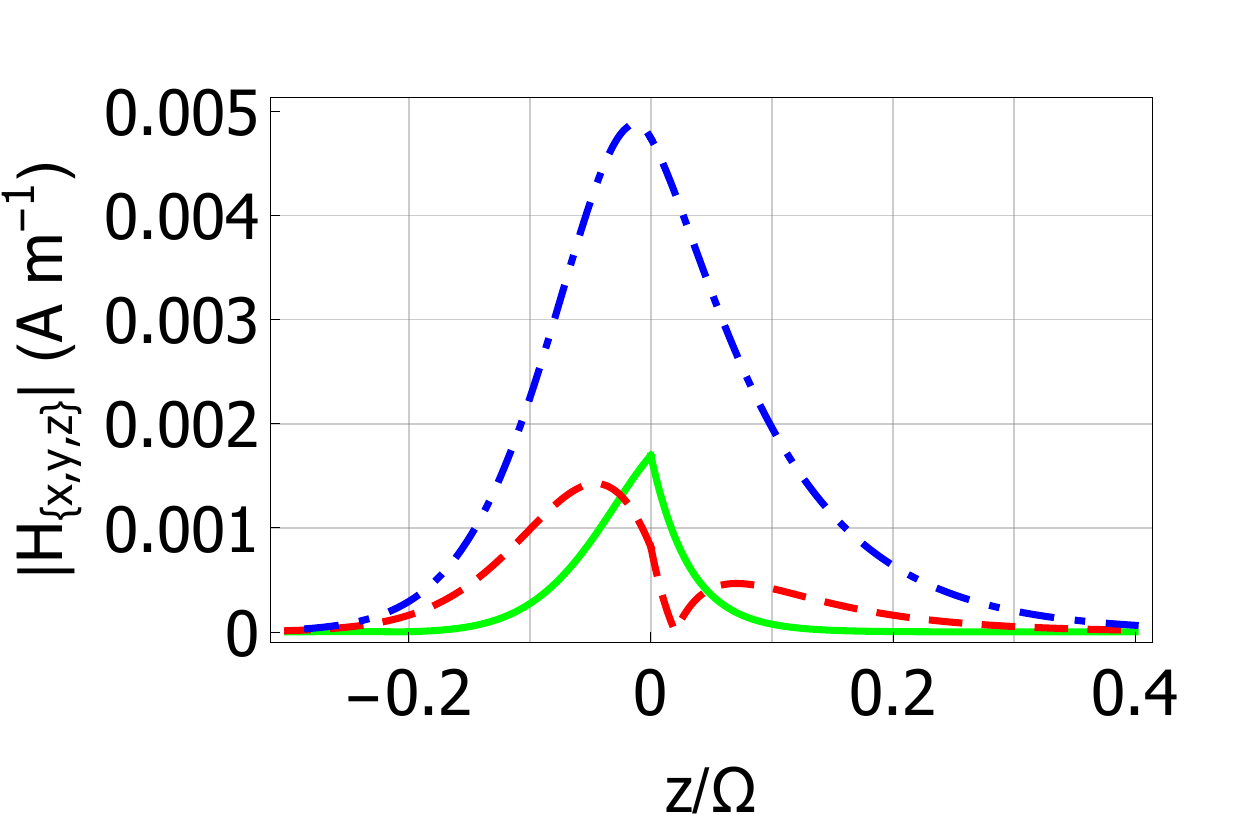}  \includegraphics[width=7.3cm]{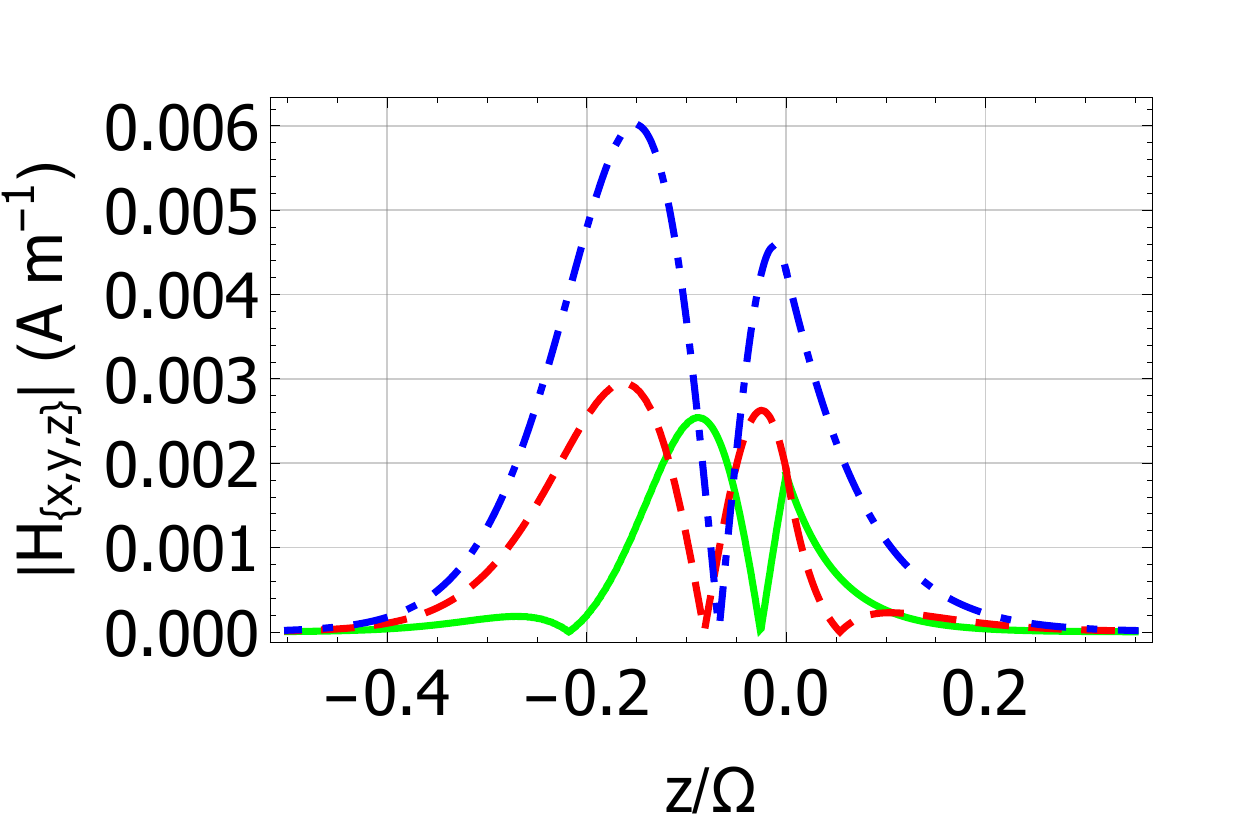} \\
 \includegraphics[width=7.3cm]{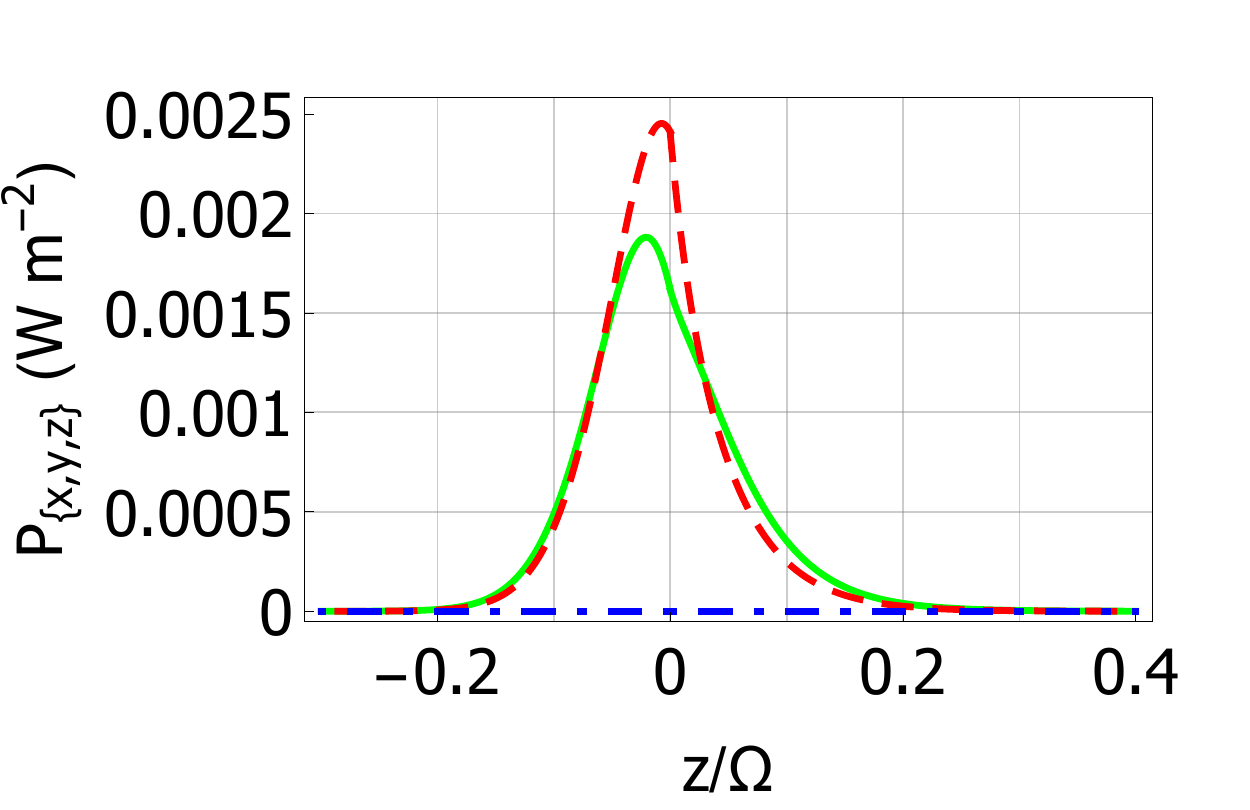}  \includegraphics[width=7.3cm]{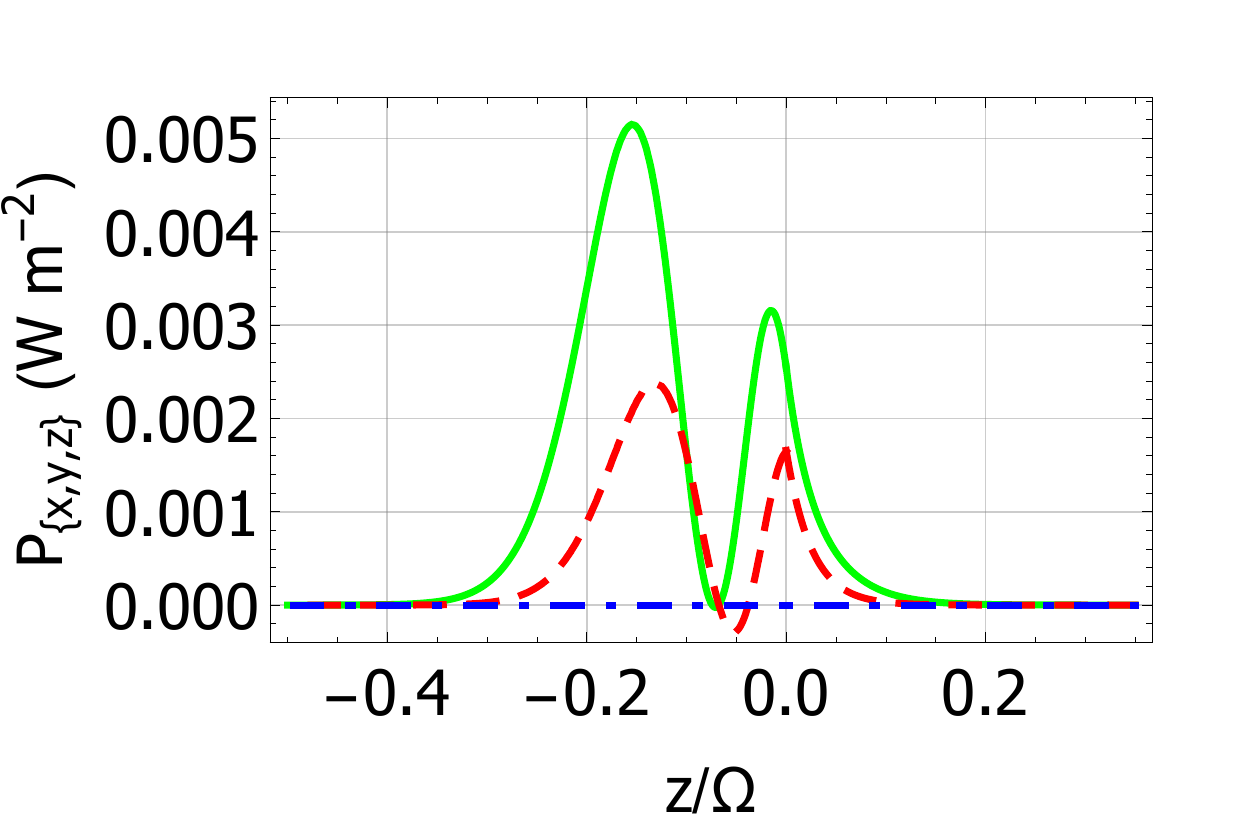}  \\
  \caption{\label{Fig3} Field profiles for (left) a  DT surface wave and (right) a DTV surface wave. Components of the quantities 
 $|\underline{E} (z\hat{\underline{u}}_{\,z}) \. \#n|$,  $|\underline{H}  (z\hat{\underline{u}}_{\,z}) \. \#n|$, and $\underline{P} (z\hat{\underline{u}}_{\,z}) \. \#n$ are plotted versus $z/\Omega$, 
 for
  $\eps^s_\calA = 2.2$, $\eps^t_\calA = 4$,  and $\chi = 0^\circ$ with  $C_{\mathcal{B}3} = 1$ V m${}^{-1}$. Left:  $\psi = 53^\circ$, $q=1.786 \, \ko$; Right: $\psi = 22.993^\circ$ , $q=1.6198 \, \ko$.
 Key:   $\#n = \ux$ green solid curves; $\#n = \uy$ red dashed  curves; $\#n = \uz$ blue broken-dashed curves.
    }
\end{figure}

The nature of the surface-wave solutions presented in  Fig.~\ref{Fig2} is further illuminated in Fig.~\ref{Fig3}
wherein
spatial profiles of the magnitudes of the Cartesian components of the electric and magnetic field phasors are provided for a DT surface wave and a DTV surface wave. As representative examples, 
 $\psi = 53^\circ$ and $q=1.786 \, \ko$ 
 were selected for the DT surface wave,
 whereas $\psi = 22.993^\circ$ and $q=1.6198 \, \ko$ were selected for the DTV surface wave.
In both cases, for $z \lessapprox - 0.2 \, \Omega$ and  $z \gtrapprox 0.05\,\Omega$, the  magnitudes of the components of the electric and magnetic field phasors  displayed in Fig.~\ref{Fig3}  decay exponentially as
  the distance $\vert{z}\vert$ from the interface plane increases. The rates of decay  in material $\calA$ and
 material $\calB$ are similar. Hence, it may be inferred
  that  the  linear decay in Eq.~\r{DV_gen_sol} 
  is dominated by the exponential decay for $z \gtrapprox 0.05\,\Omega$.

Insight in to the localization of the  surface waves is also provided by profiles of 
  the Cartesian components  of the time-averaged Poynting vector
\begin{equation}
\underline{P}  (\#r) = \frac{1}{2} \mbox{Re} \lec \, \underline{E}  (\#r)  \times 
\underline{H}^*  (\#r) \, \ric
\end{equation}
that are presented
in Fig.~\ref{Fig3}.
 These profiles show that
energy flow for both the DT and the DTV surface waves is concentrated in directions parallel to the interface plane $z=0$. Furthermore, the energy densities of the surface waves are concentrated not at the interface $z=0$, but at a distance of approximately $0.15 \, \Omega$ from the interface in material $\calB$ for the DTV wave, and a 
distance of approximately $0.02 \, \Omega$ from the interface in material $\calB$ for the DT wave.

\begin{figure}[!htb]

\centering
 \includegraphics[width=7.3cm]{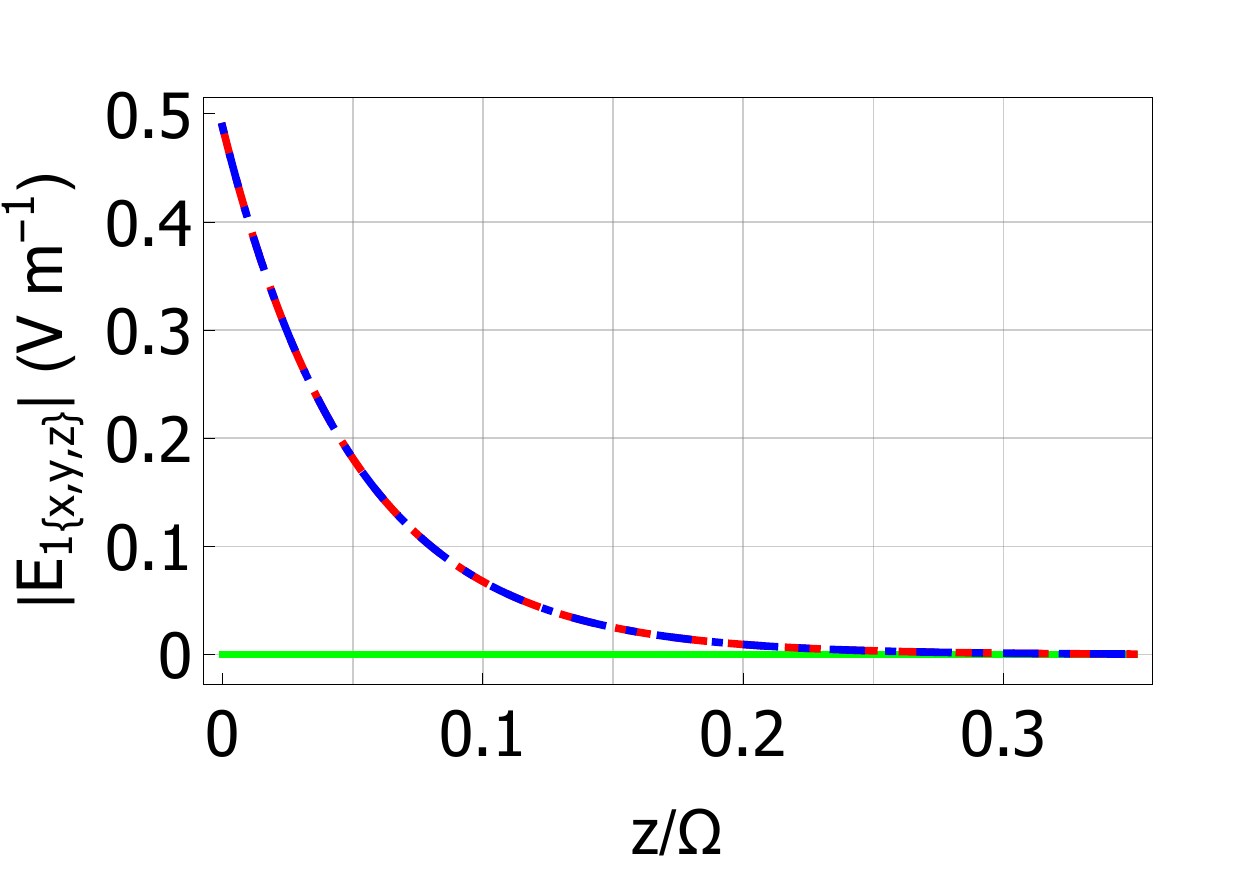}  \includegraphics[width=7.3cm]{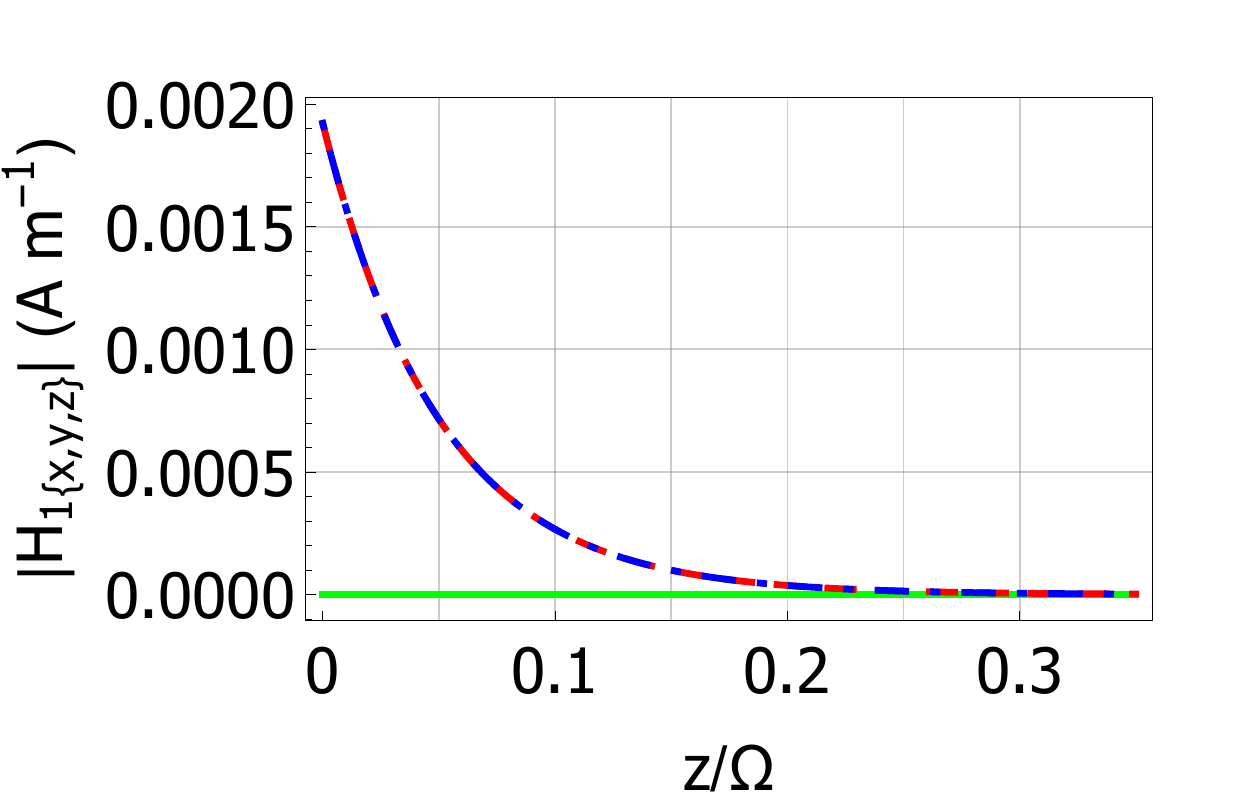}  \\
 \includegraphics[width=7.3cm]{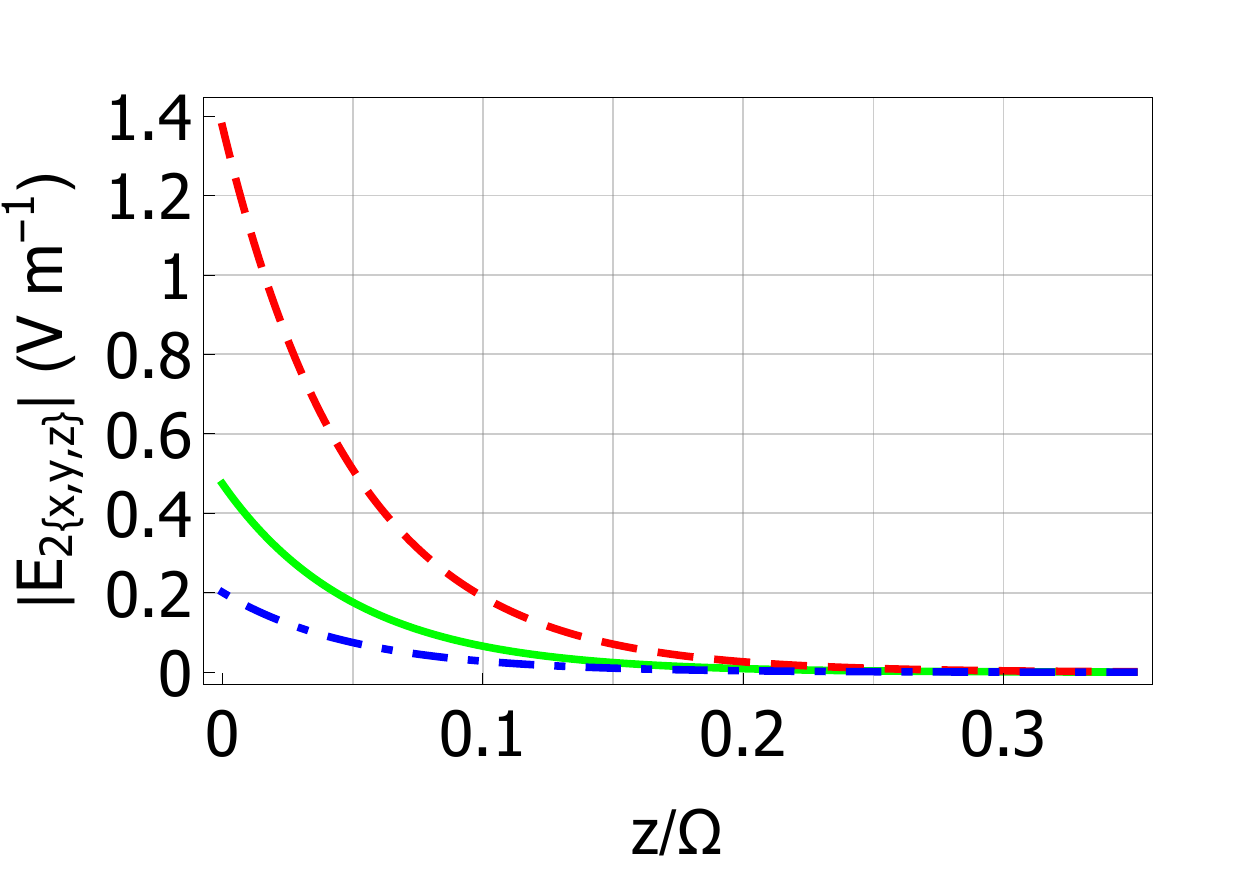}  \includegraphics[width=7.3cm]{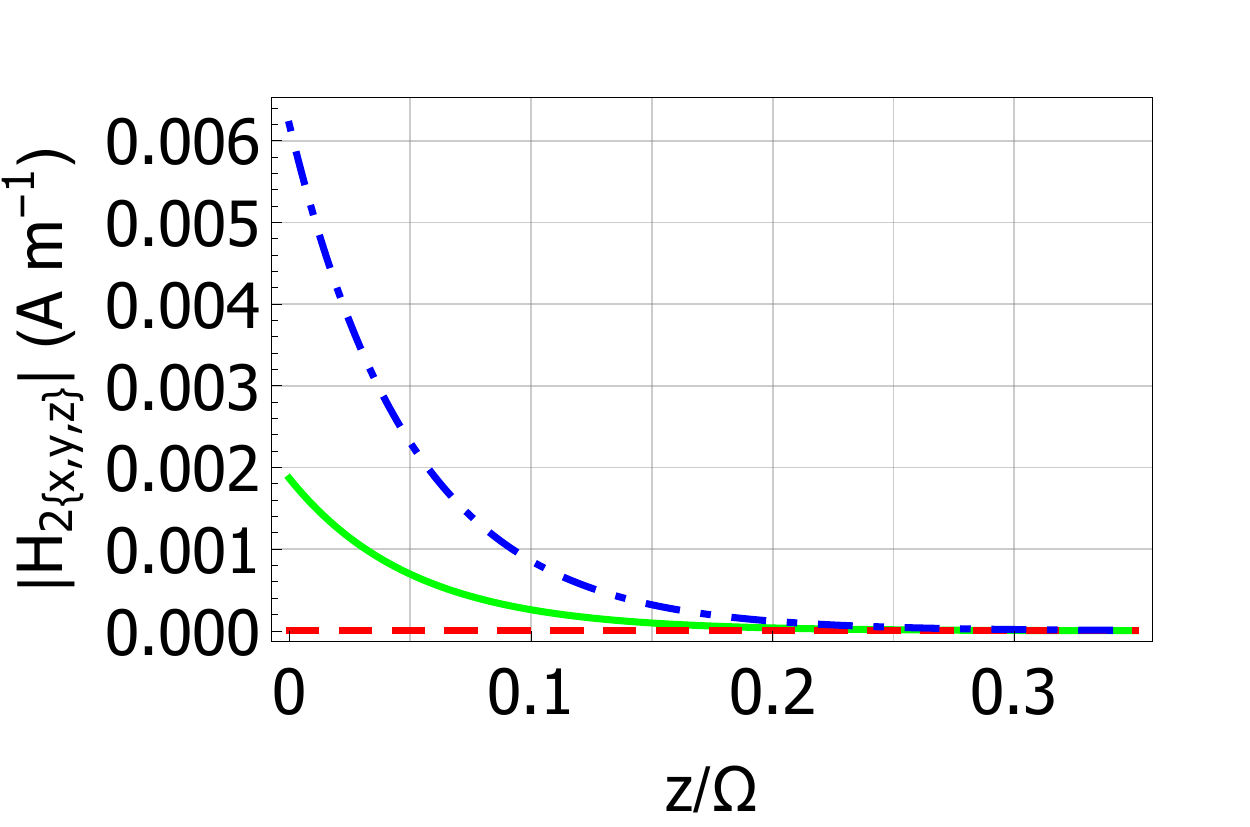} \\
 \includegraphics[width=7.3cm]{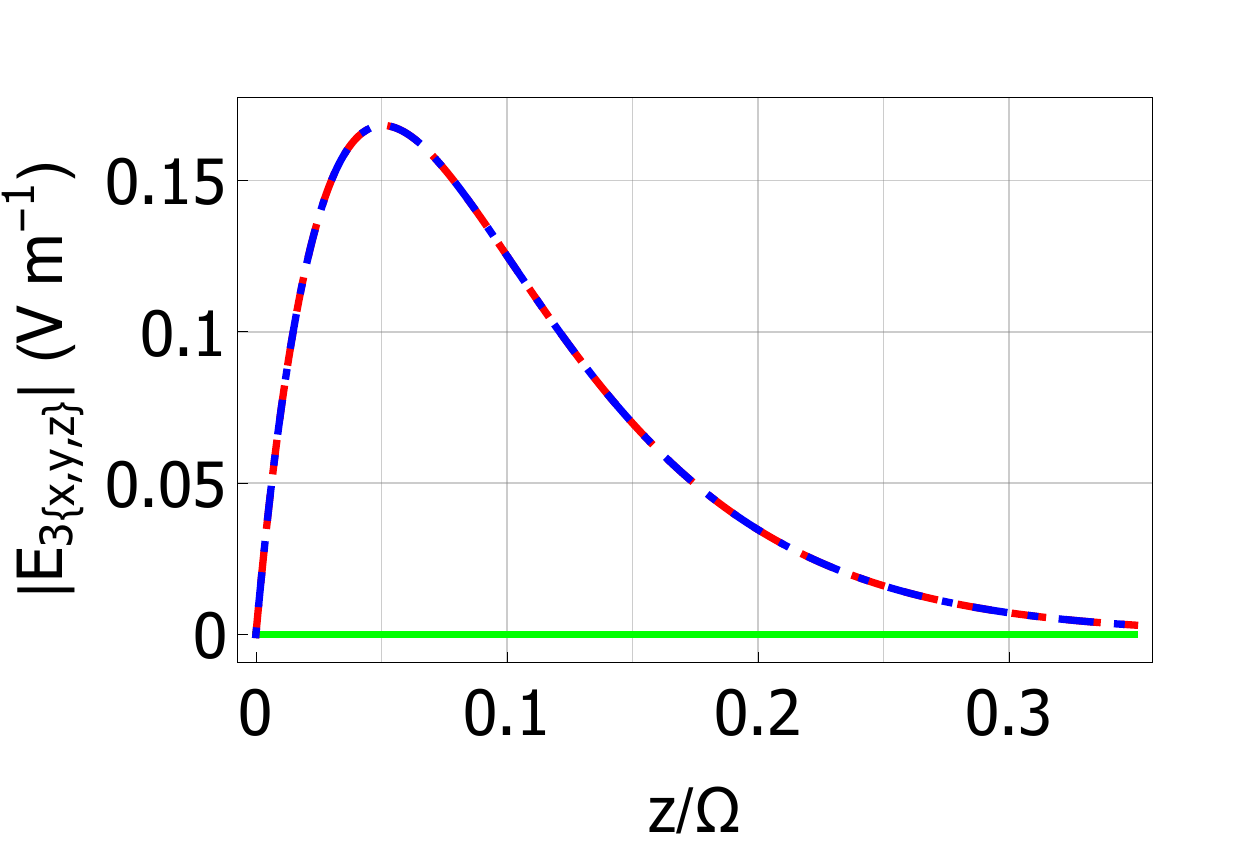}  \includegraphics[width=7.3cm]{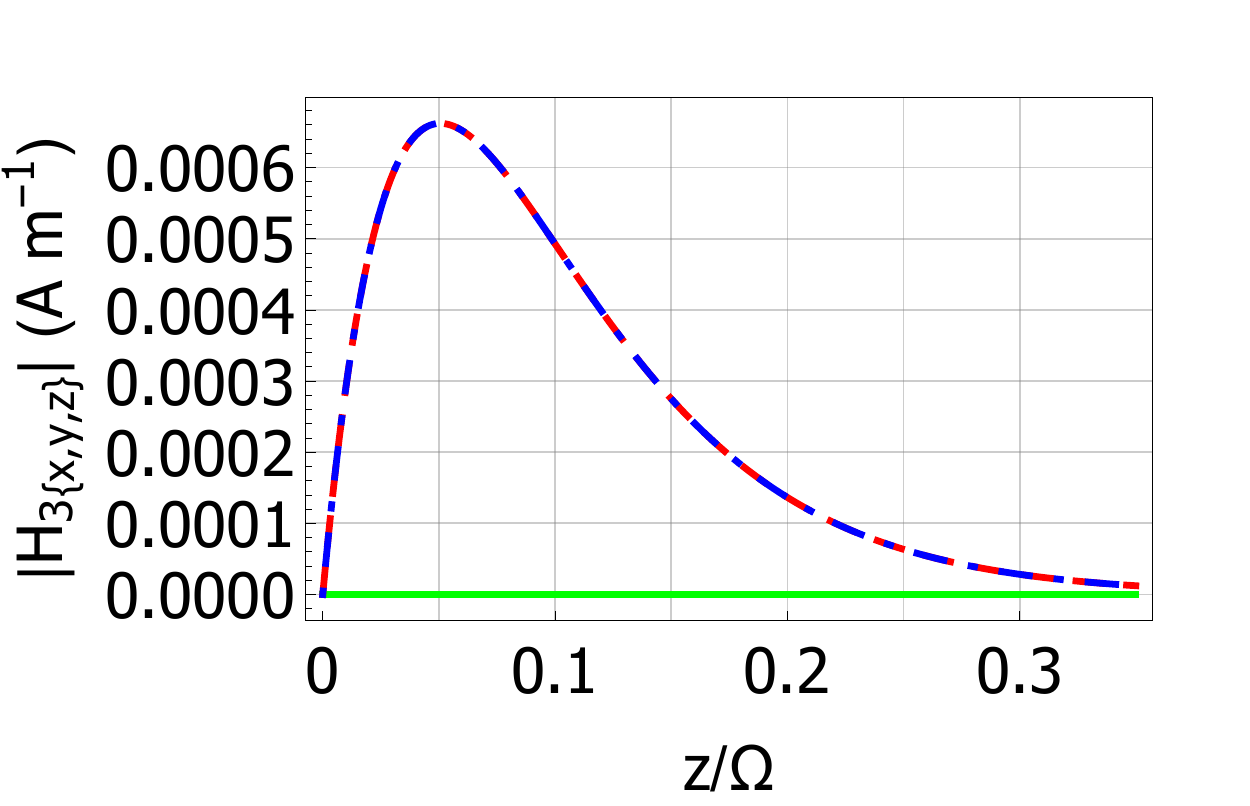}  \\
  \caption{\label{Fig4} Field profiles for
  the contributions to the DTV surface-wave solution represented in Fig.~\ref{Fig3} (right),  per Eq.~\r{f_contributions}.
   Components of the quantities 
 $|\underline{E}_\ell (z\hat{\underline{u}}_{\,z}) \. \#n|$ and  $|\underline{H}_\ell  (z\hat{\underline{u}}_{\,z}) \. \#n|$,
 where $\ell \in \lec 1, 2, 3 \ric$,
  are plotted versus $z/\Omega$, 
 for
  $\eps^s_\calA = 2.2$, $\eps^t_\calA = 4$,  and $\chi = 0^\circ$ with  $C_{\mathcal{B}3} = 1$ V m${}^{-1}$, with $\psi = 22.993^\circ$ and  $q=1.6198 \, \ko$.
 Key:   $\#n = \ux$ green solid curves; $\#n = \uy$ red dashed  curves; $\#n = \uz$ blue broken-dashed curves.
    }
\end{figure}

Let us consider further the anatomy of the DTV surface-wave solution, as provided in  Eq.~\r{DV_gen_sol}
for $z>0$. Three contributions to $\les\#f  (z)\ris$ may be identified, namely
\begin{equation} \l{f_contributions}
\les\#f  (z)\ris =
\les\#f_1  (z)\ris + \les\#f_2  (z)\ris +\les\#f_3  (z)\ris, \qquad z>0,
\end{equation}
wherein
\begin{equation}
\left.
\begin{array}{l}
\les\#f_1  (z)\ris =  C_{\calA 1}  \les\#v_{\calA }\ris \, \exp \le i \alpha_{\calA } z \ri \vspace{6pt}\\
\les\#f_2  (z)\ris =  C_{\calA 2}  \les\#w_{\calA }\ris \, \exp \le i \alpha_{\calA } z \ri \vspace{6pt}\\
\les\#f_3  (z)\ris =  C_{\calA 2} \lec i z  \les\#v_{\calA }\ris \ric \, \exp \le i \alpha_{\calA } z \ri 
\end{array}
\right\}.
\end{equation}
The $x$ and $y$ components of the electric and magnetic field phasors are assembled to form the 4-vectors $\les\#f_\ell  (z)\ris $ for $\ell = 1, 2,$ and 3, per Eq.~\r{f-def}. The corresponding $z$ components are delivered by means of
Eqs.~\r {z_comp}. Profiles of the magnitudes of the  Cartesian components of the 
electric and magnetic field phasors comprising  $\les\#f_\ell  (z)\ris $, $\ell \in \lec 1, 2, 3 \ric$, are plotted 
for $z>0$ 
in
Fig.~\ref{Fig4},
for the DTV surface-wave solution represented in Fig.~\ref{Fig3}.
Close to the planar interface, i.e., for $z <  0.05 \Omega$, 
the magnitudes presented in Fig.~\ref{Fig4} corresponding to the  exponentially decaying contributions $\les\#f_1  (z)\ris$ and $\les\#f_2  (z)\ris$ are much larger than the magnitudes corresponding to the mixed linear-exponential contribution $\les\#f_3  (z)\ris$.
By comparing with the profiles of 
$|\underline{E} (z\hat{\underline{u}}_{\,z}) \. \#n|$ and  $|\underline{H}  (z\hat{\underline{u}}_{\,z}) \. \#n|$ with $\#n \in \lec \ux, \uy, \uz \ric$ in Fig.~\ref{Fig3}  for $z>0$, we infer that the  mixed linear-exponential contribution $\les\#f_3  (z)\ris$ has a stronger effect  on $|\underline{E} (z\hat{\underline{u}}_{\,z}) \. \uz|$
than on $|\underline{E} (z\hat{\underline{u}}_{\,z}) \. \ux|$ and $|\underline{E} (z\hat{\underline{u}}_{\,z}) \. \uy|$, and a stronger effect on 
$|\underline{H} (z\hat{\underline{u}}_{\,z}) \. \uy|$
than on $|\underline{H} (z\hat{\underline{u}}_{\,z}) \. \ux|$ and $|\underline{H} (z\hat{\underline{u}}_{\,z}) \. \uz|$.

\begin{figure}[!htb]
\centering
\begin{subfigure}{0.49\textwidth}
 \includegraphics[width=7.3cm]{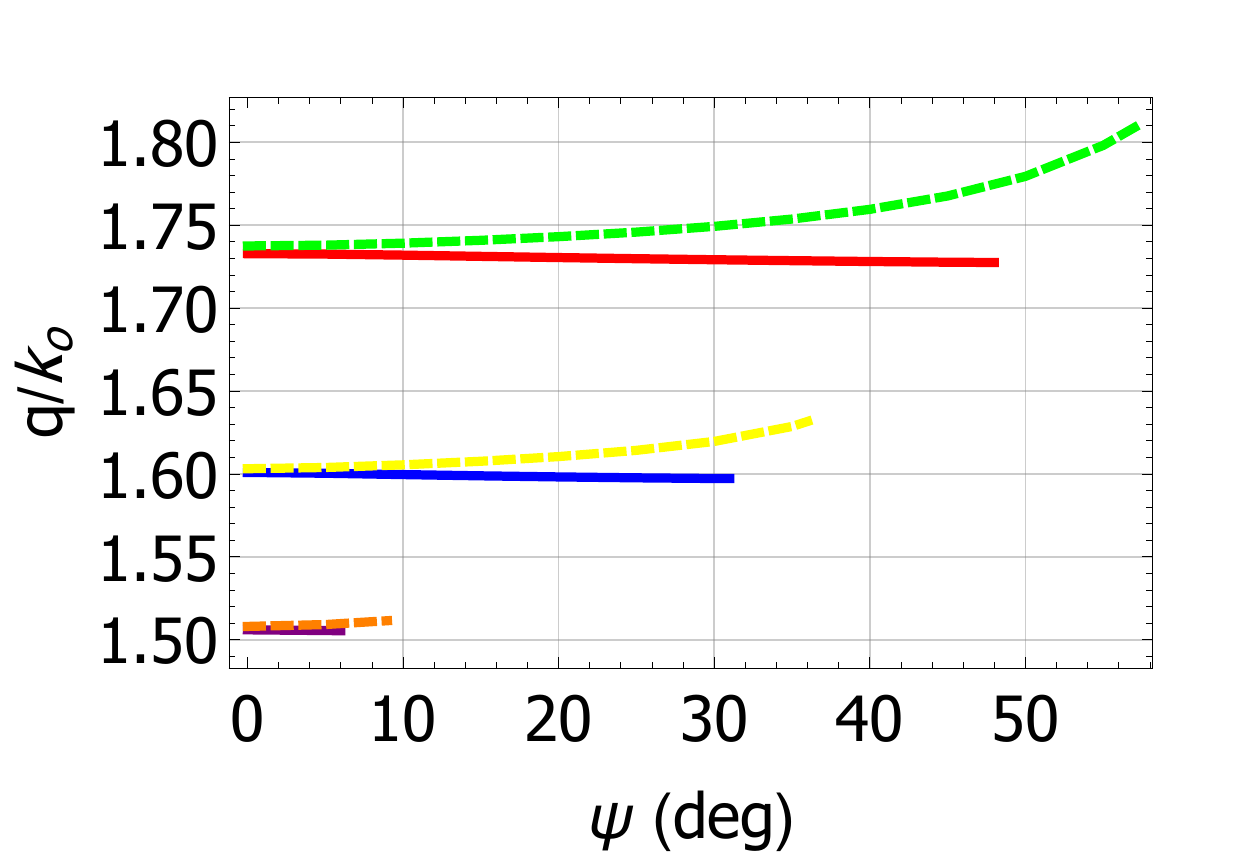} 
 \caption{}
 \end{subfigure} \hfill
 \begin{subfigure}{0.49\textwidth}
  \includegraphics[width=7.3cm]{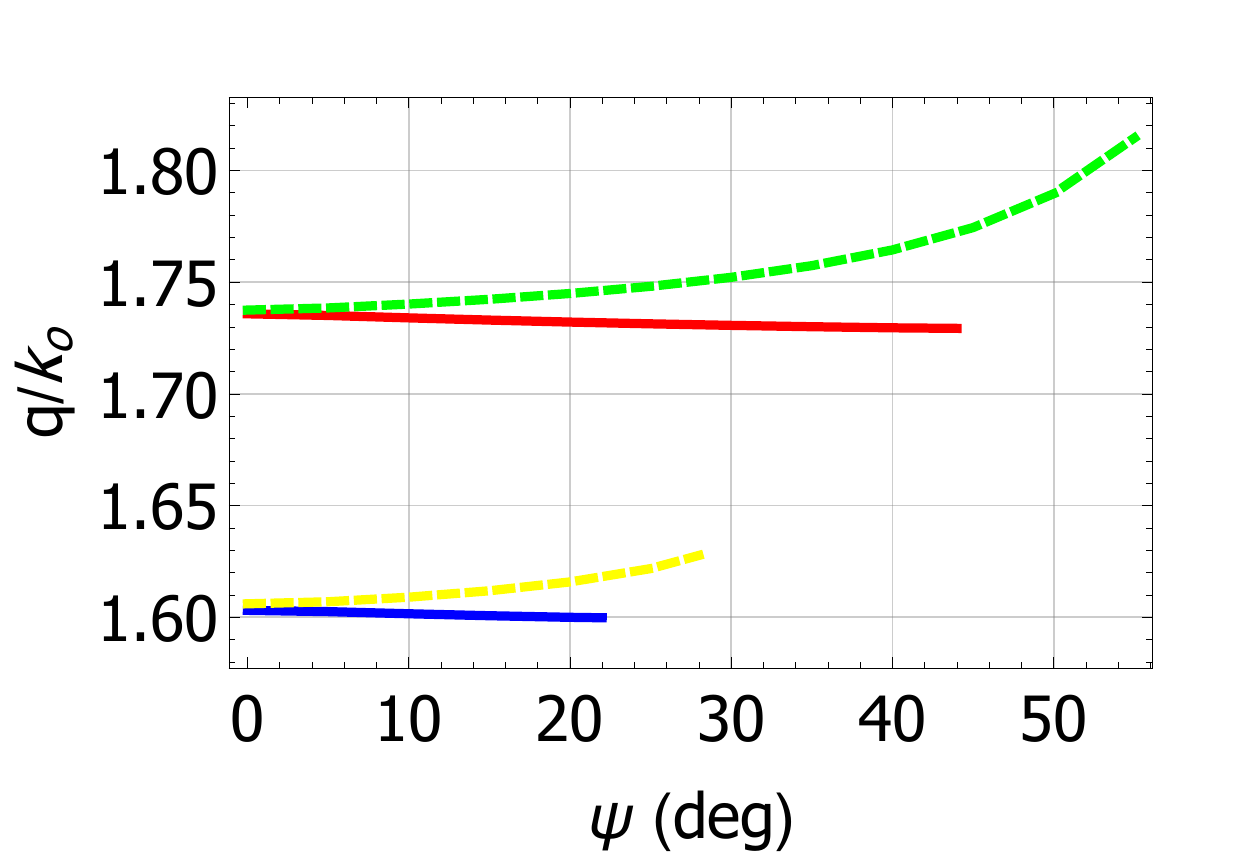}
 \caption{}
  \end{subfigure}\\
  \begin{subfigure}{0.49\textwidth}
     \includegraphics[width=7.3cm]{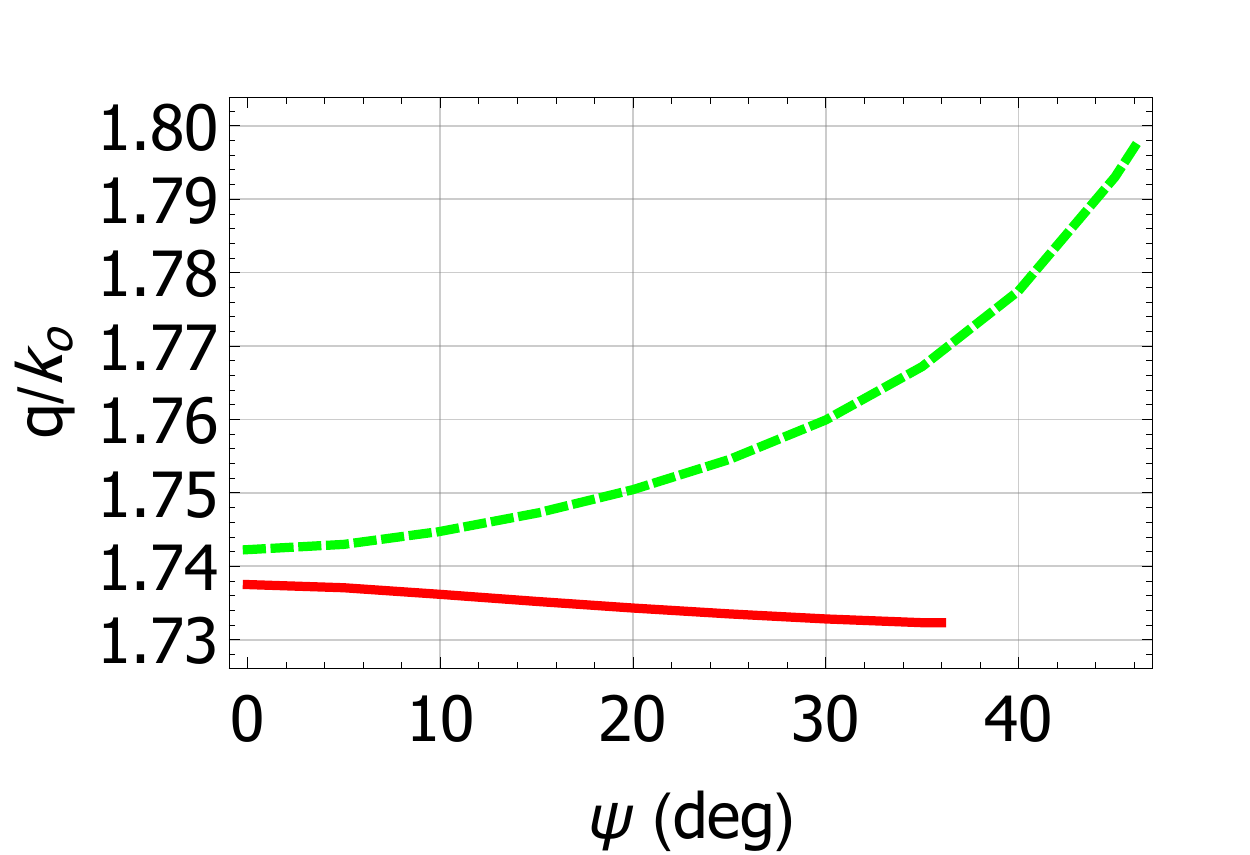} 
     \caption{}
     \end{subfigure}
 \caption{\label{Fig5} 
 Plots of  $q/\ko $
versus 
$\psi $ 
for  $\gamma=1$ and $\eps^s_\calA=2.2$, when (a)  $\chi=10^\circ$,  (b) $\chi=20^\circ$, and  (c) $\chi=30^\circ$. 
The  curves represent DT surface-wave solutions: there are 6 branches for (a), 
4  for (b), and 2   for (c).
    }
\end{figure}

The influence of the orientation of the optic axis of material $\calA$ is taken up in Figs.~\ref{Fig5}(a-c), wherein
 plots of  $q/\ko $
versus 
$\psi $ are provided for
 (a)  $\chi=10^\circ$,  (b) $\chi=20^\circ$, and  (c) $\chi=30^\circ$. 
 For these calculations,   $\gamma=1$ and $\eps^s_\calA=2.2$. The DT surface-wave
 solutions are organized as 6 branches for $\chi=10^\circ$, 
4 branches for $\chi=20^\circ$, and 2 branches   for $\chi=30^\circ$. 
The characteristics of the $q/\ko $ vs.
$\psi $ curves in Figs.~\ref{Fig5} and  \ref{Fig2} are quite similar.
The number of DT branches   decreases as $\chi$ increases, with no DT surface-wave solutions at all being found for $\chi > 45^\circ$.

Not a single DTV surface-wave solution exists in  Fig.~\ref{Fig5}. Indeed, no DTV surface-wave solution was found
by us
for $\chi > 0^\circ$. An analogous  result holds for  Dyakonov--Voigt surface waves \c{ZML_JOSAB}.

\begin{figure}[!htb]
\centering
 \includegraphics[width=7.3cm]{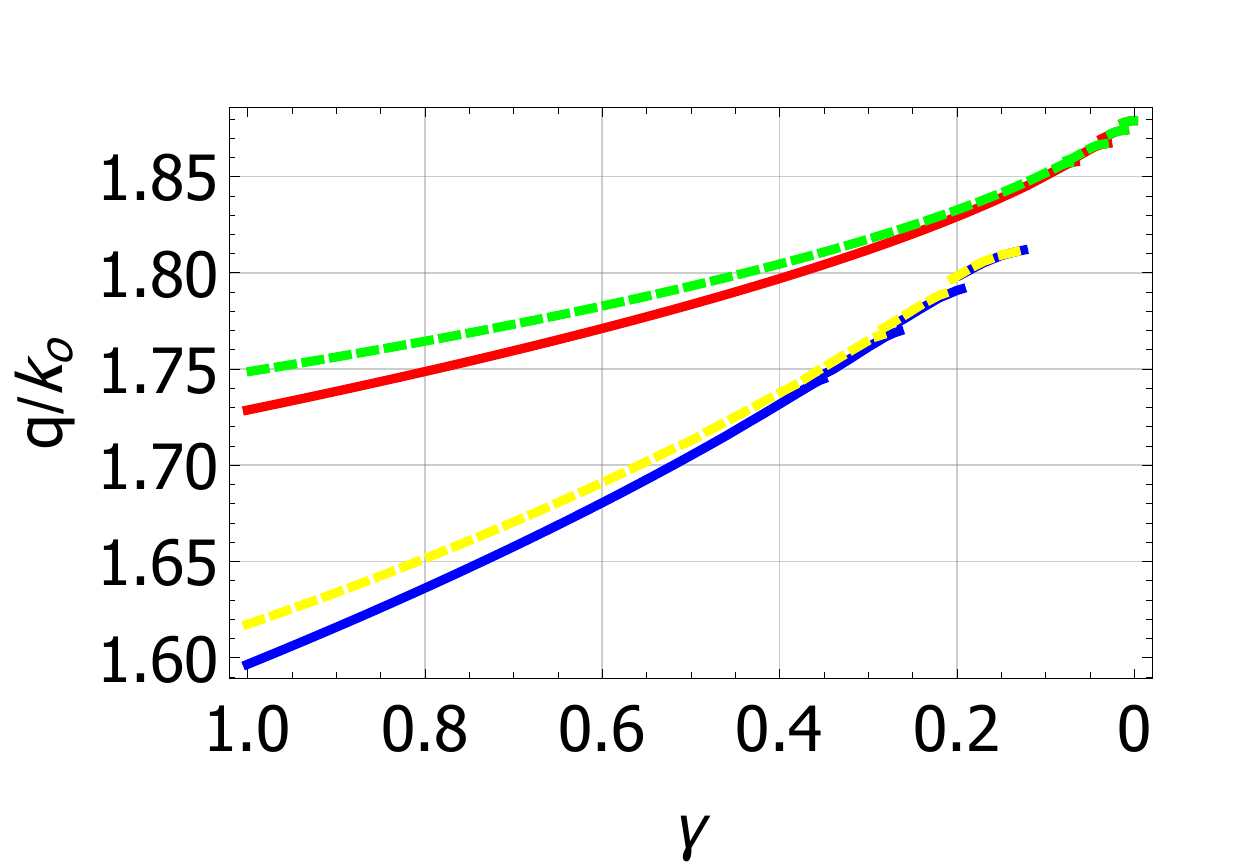} 
\hfill
  \includegraphics[width=7.3cm]{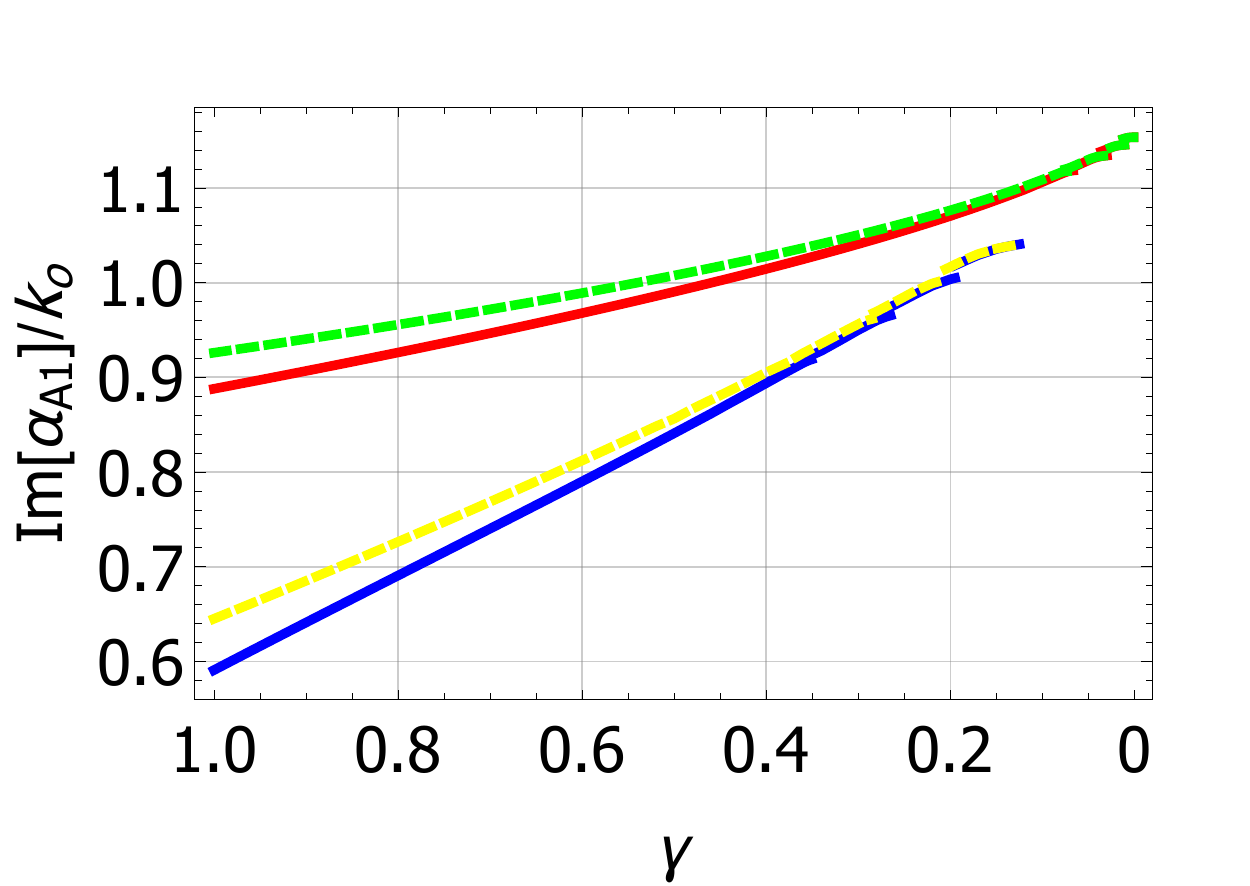}
\\
     \includegraphics[width=7.3cm]{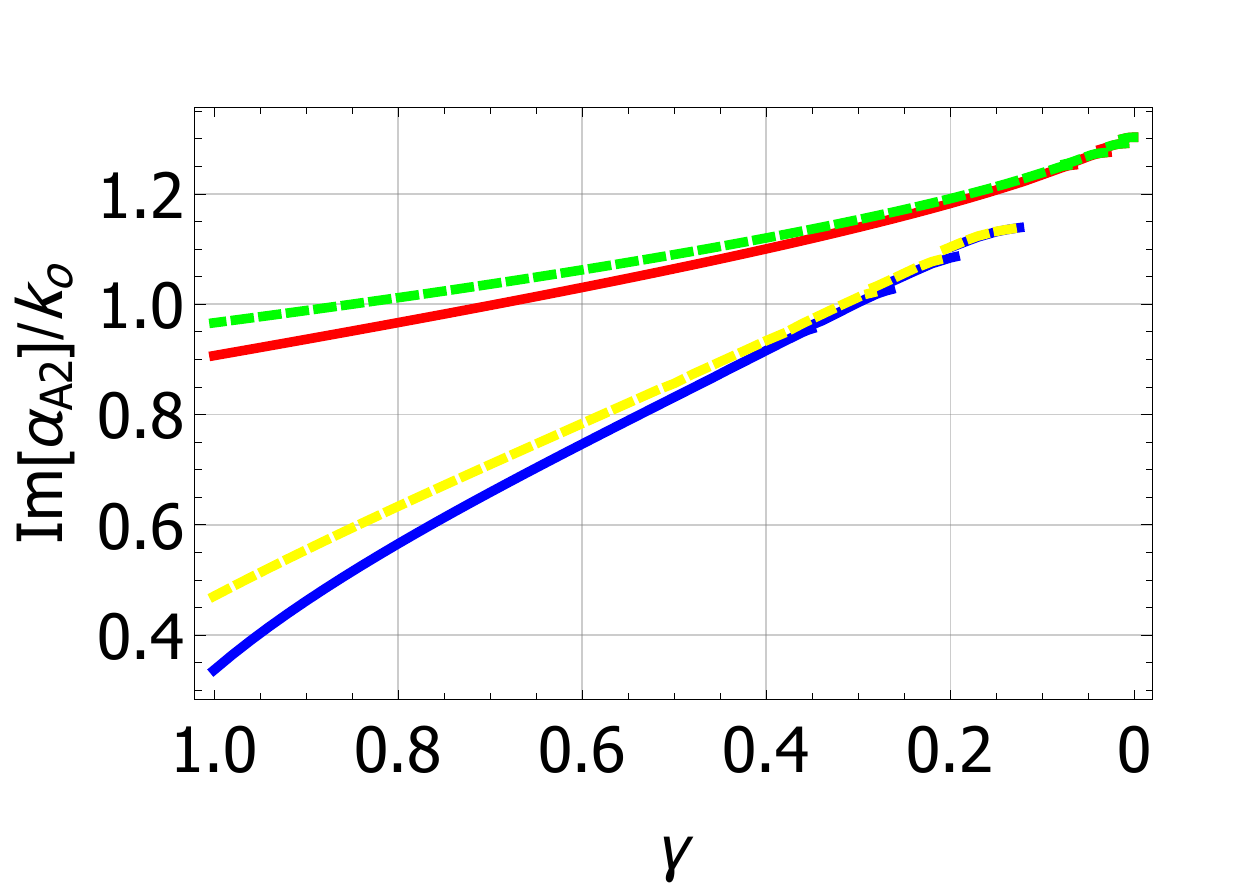} 
     \hfill
  \includegraphics[width=7.3cm]{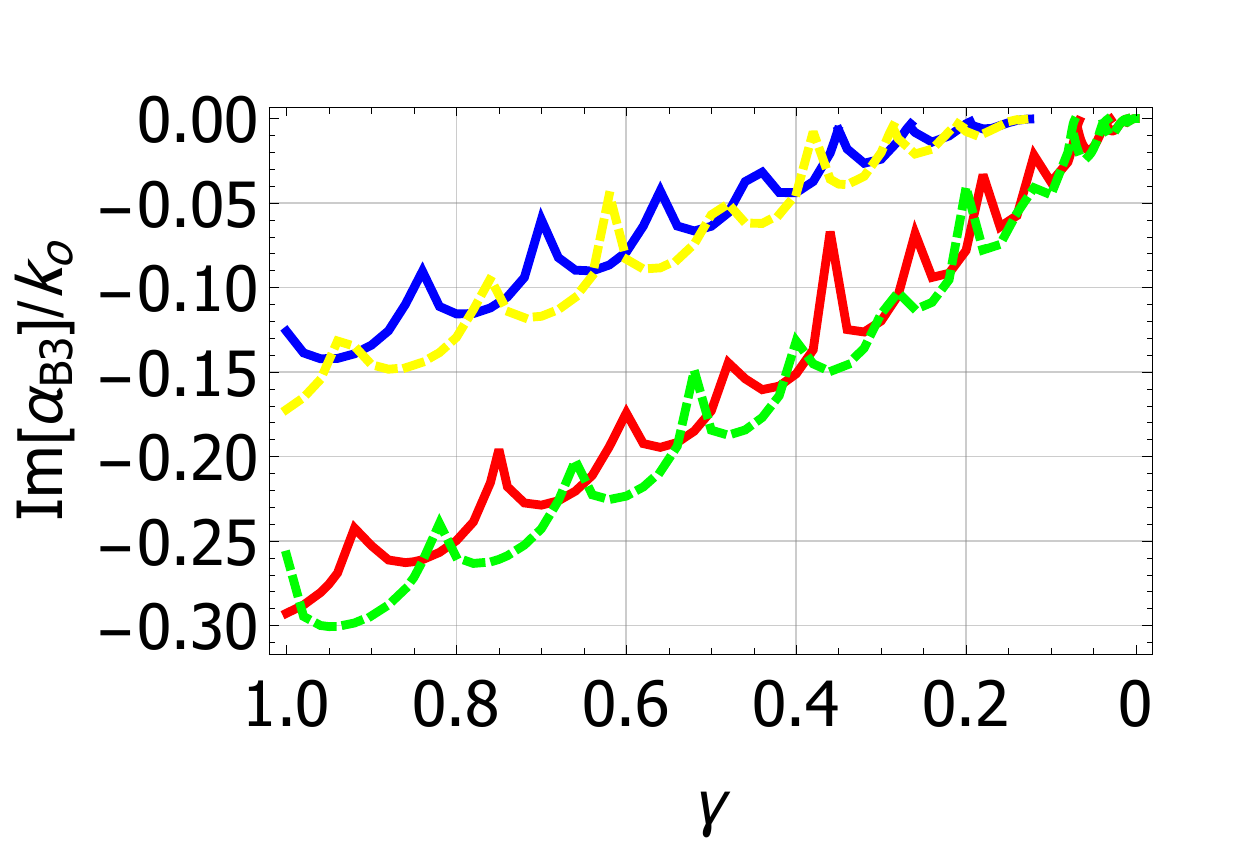}
\\
     \includegraphics[width=7.3cm]{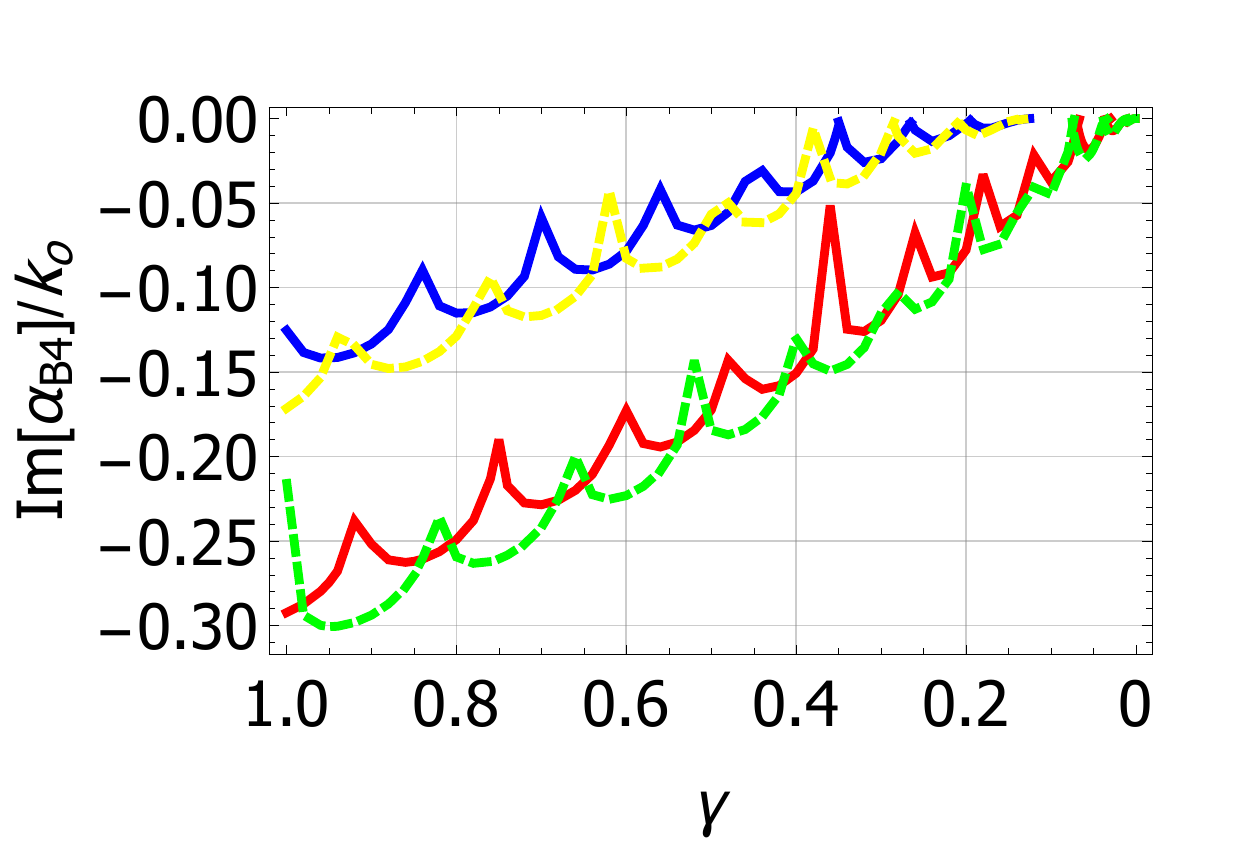} 
 \caption{\label{Fig6} 
 Plots of  $q/\ko $, $\mbox{Im} \lec \alpha_{\calA1} \ric/\ko$, $\mbox{Im} \lec \alpha_{\calA2} \ric/\ko$, 
$\mbox{Im} \lec \alpha_{\calB3} \ric/\ko$, and $\mbox{Im} \lec \alpha_{\calB4} \ric/\ko$
versus 
 $\gamma$, for $\psi = 30^\circ$, $\eps^s_\calA=2.2$, and $\chi=0^\circ$.
 The  curves represent DT surface-wave solutions: there are 4 branches for  $0.12 < \gamma \leq 1 $ and 2 branches   for  $0<\gamma < 0.12$.
    }
\end{figure}

\begin{figure}[!htb]
\centering
 \includegraphics[width=7.3cm]{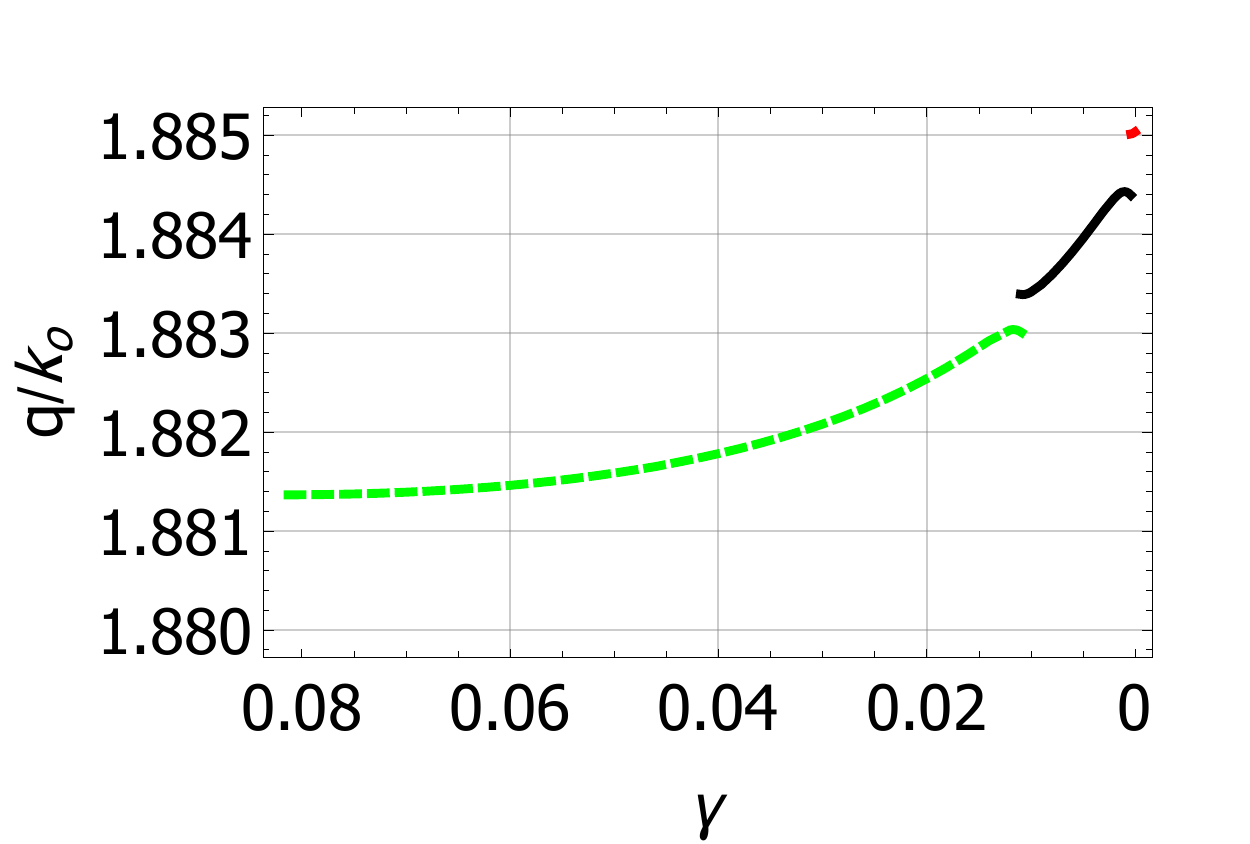} 
\hfill
  \includegraphics[width=7.3cm]{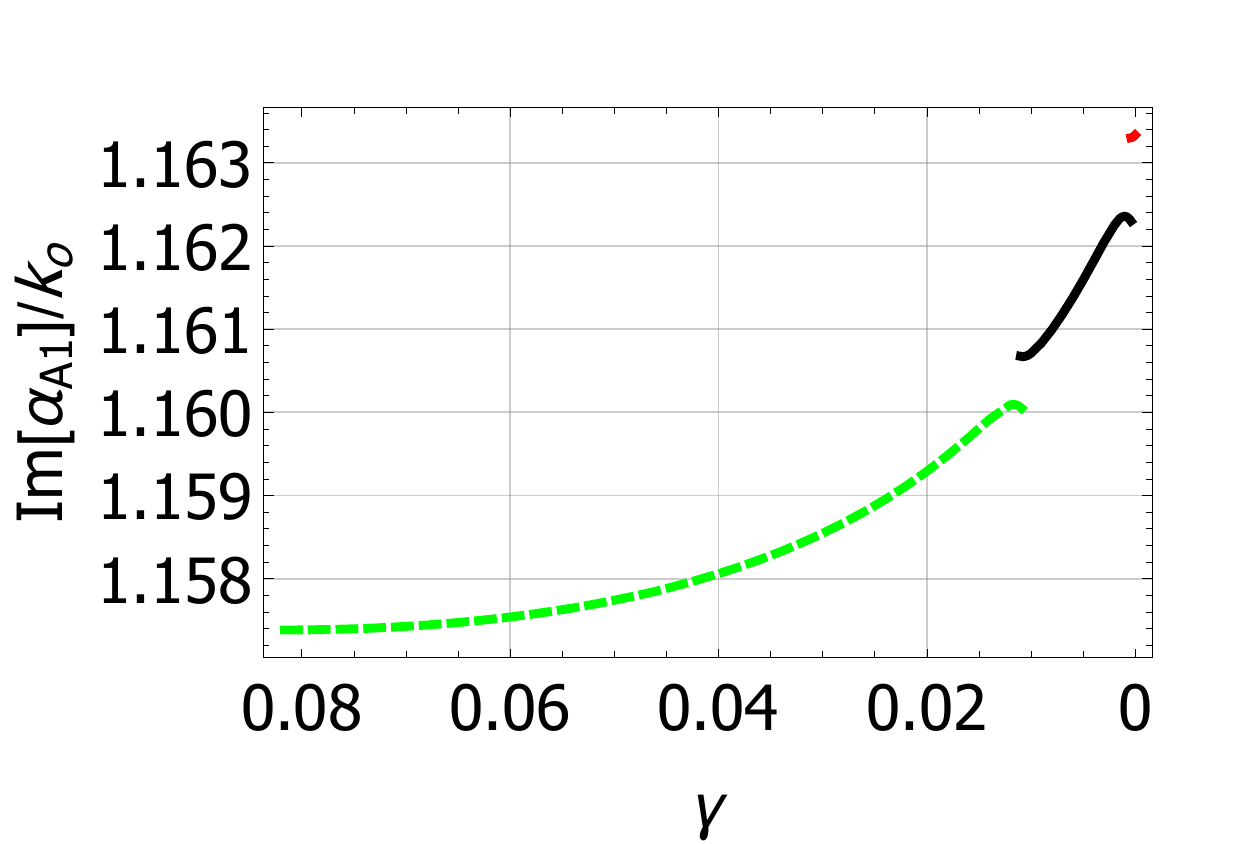}
\\
     \includegraphics[width=7.3cm]{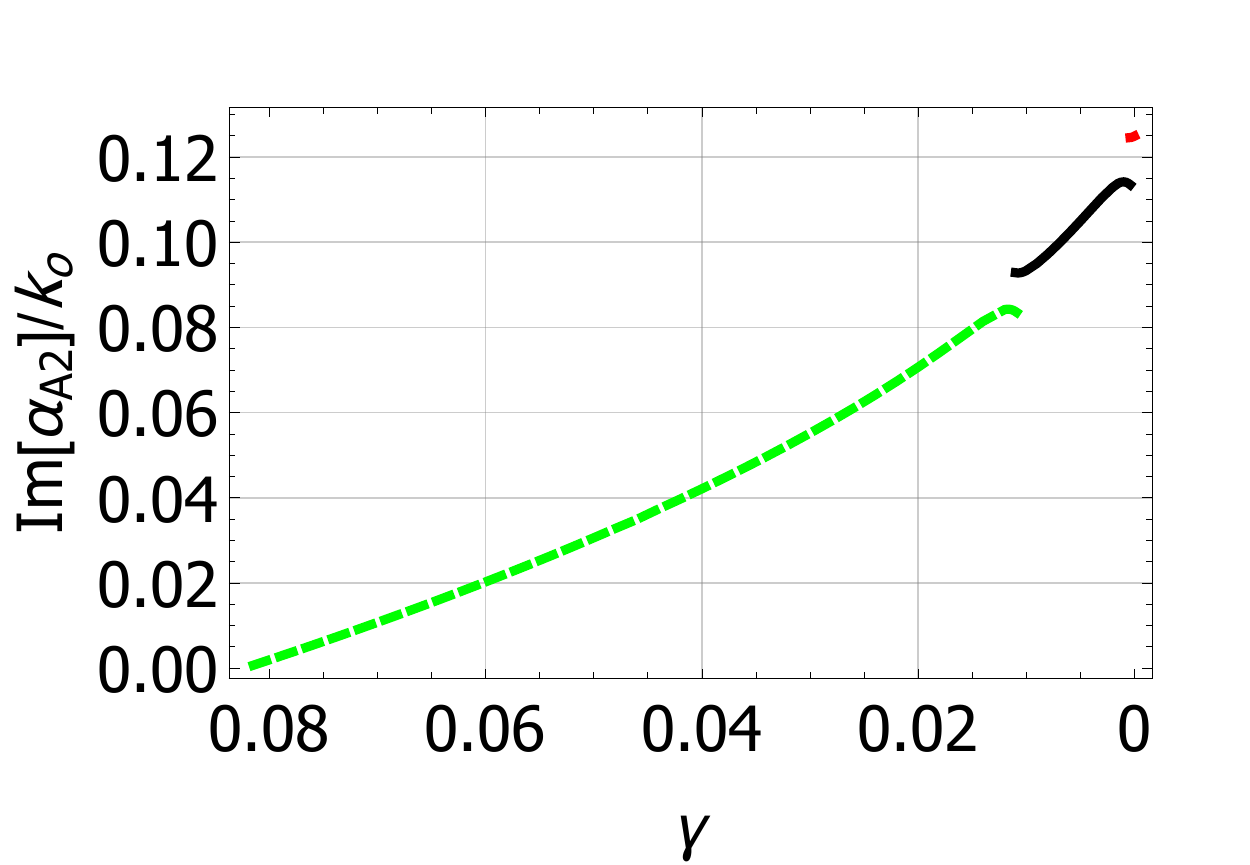} 
     \hfill
  \includegraphics[width=7.3cm]{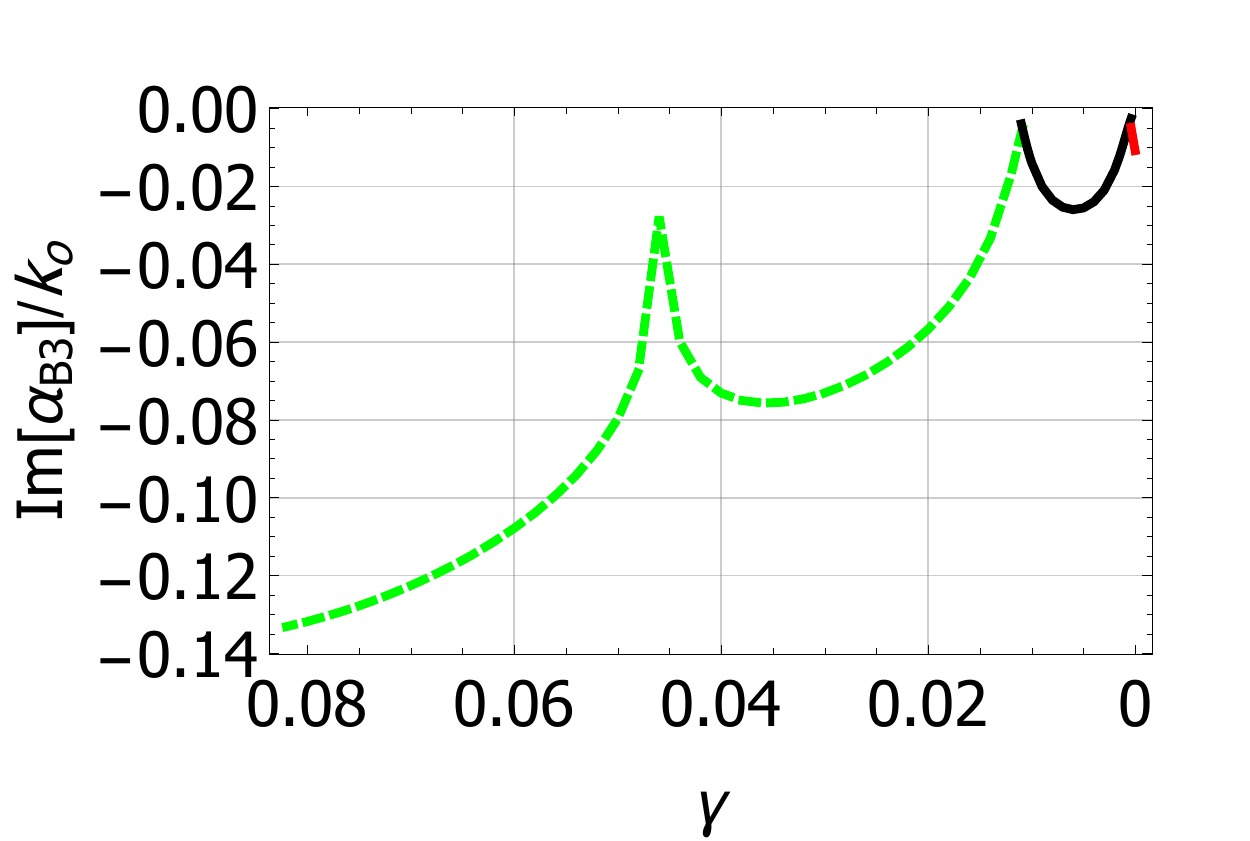}
\\
     \includegraphics[width=7.3cm]{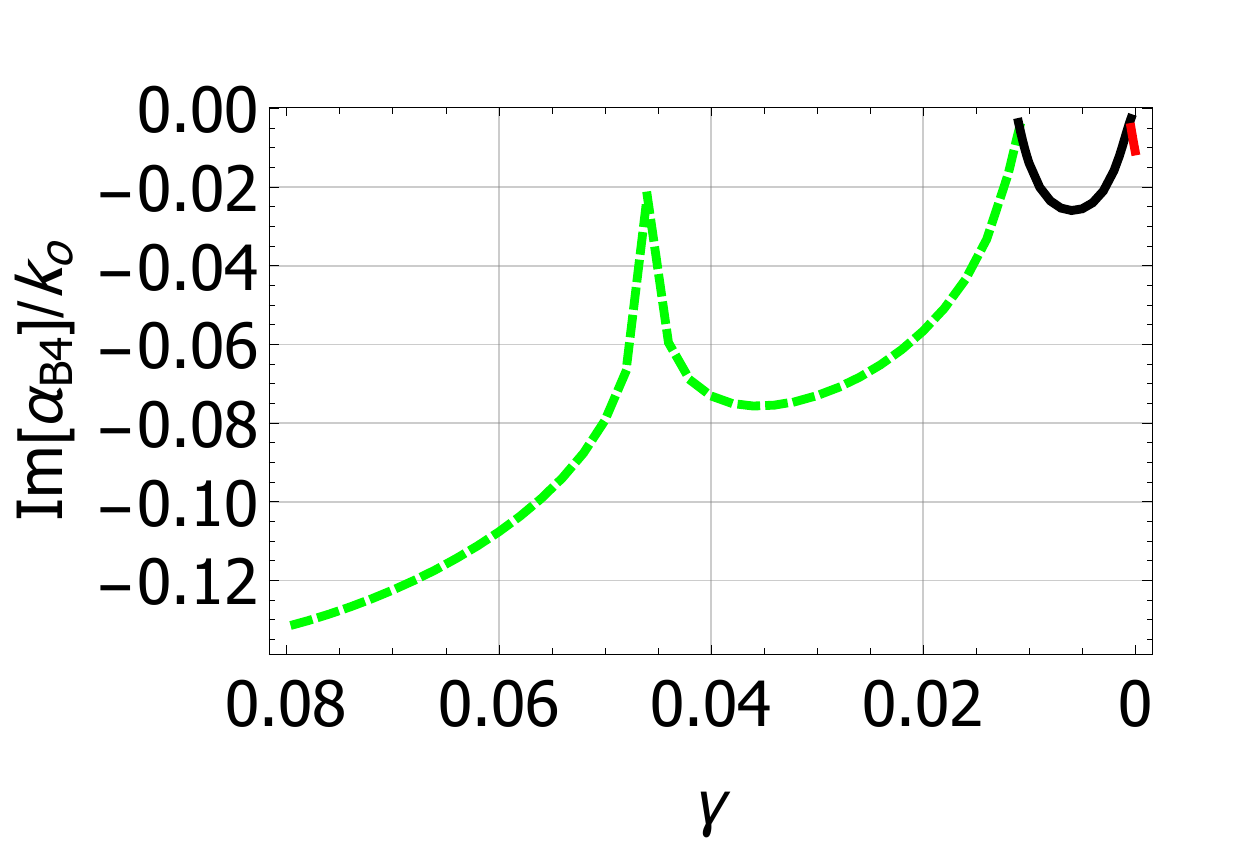} 
 \caption{\label{Fig7} 
 Plots of  $q/\ko $, $\mbox{Im} \lec \alpha_{\calA1} \ric/\ko$, $\mbox{Im} \lec \alpha_{\calA2} \ric/\ko$, 
$\mbox{Im} \lec \alpha_{\calB3} \ric/\ko$, and $\mbox{Im} \lec \alpha_{\calB4} \ric/\ko$ 
versus 
 $\gamma$, for $\psi = 66.5^\circ$, $\eps^s_\calA=2.2$, and $\chi=0^\circ$.
The  curves represent DT surface-wave solutions: there is only one branch  for   $\gamma < 0.082$.
    }
\end{figure}

The influence of the  amplitude  of periodic variation in  dielectric properties along the negative $z$ axis is taken up in Figs.~\ref{Fig6} and \ref{Fig7}.
 Plots of  $q/\ko $, $\mbox{Im} \lec \alpha_{\calA1} \ric/\ko$, $\mbox{Im} \lec \alpha_{\calA2} \ric/\ko$, 
$\mbox{Im} \lec \alpha_{\calB3} \ric/\ko$, and $\mbox{Im} \lec \alpha_{\calB4} \ric/\ko$, 
versus 
 $\gamma$ are presented in Fig.~\ref{Fig6} for $\psi = 30^\circ$, $\chi=0^\circ$, and $\eps^s_\calA=2.2$.
There are 4 branches of DT surface-wave solutions for $0.12 < \gamma \leq 1 $ and 2 branches for  $0<\gamma < 0.12$.
The quantities  $q/\ko $, $\mbox{Im} \lec \alpha_{\calA1} \ric/\ko$, and $\mbox{Im} \lec \alpha_{\calA2} \ric/\ko$ all  increase uniformly as $\gamma$ decreases, whereas 
$\mbox{Im} \lec \alpha_{\calB3} \ric/\ko$ and $\mbox{Im} \lec \alpha_{\calB4} \ric/\ko$ generally increase as $\gamma$ decreases. For the 2 branches that exist for $0<\gamma < 0.12$, both $\mbox{Im} \lec \alpha_{\calB3} \ric/\ko$ and $\mbox{Im} \lec \alpha_{\calB4} \ric/\ko$ become null valued in the limit as $\gamma$ approaches zero.
Accordingly, these 2 branches   do not represent surface waves in the limiting case $\gamma \to 0$  because the
conditions
$\mbox{Im} \lec \alpha_{\calB3} \ric/\ko < 0$ and $\mbox{Im} \lec \alpha_{\calB4} \ric/\ko < 0$, which must be satisfied for surface-wave propagation, are not  then satisfied.

In Fig.~\ref{Fig7} plots analogous to those for Fig.~\ref{Fig6} are provided for the case of $\psi = 66.5^\circ$. In order to better illustrate the regime in which $\gamma$ approaches zero, the plots in Fig.~\ref{Fig7} focus on the  range $\gamma \in \les 0.082, 0 \ri$. In this case there is only one   branch   of DT surface-wave solutions. No DTV surface-wave solutions exist in  $\gamma \in \les 1, 0.082 \ri$.
All the quantities plotted in Fig.~\ref{Fig7} generally increase   as $\gamma$ decreases, albeit the curves for
  $q/\ko $, $\mbox{Im} \lec \alpha_{\calA1} \ric/\ko$, and $\mbox{Im} \lec \alpha_{\calA2} \ric/\ko$  are discontinuous.
  The curves for
   $\mbox{Im} \lec \alpha_{\calA1} \ric/\ko$ and $\mbox{Im} \lec \alpha_{\calA2} \ric/\ko$
are similar, and so are the curves  for
  $\mbox{Im} \lec \alpha_{\calB3} \ric/\ko$ and $\mbox{Im} \lec \alpha_{\calB4}\ric /\ko$.
  Unlike in Fig.~\ref{Fig6},  
  $\mbox{Im} \lec \alpha_{\calB3} \ric/\ko$ and $\mbox{Im} \lec \alpha_{\calB4}\ric /\ko$
  do not become null valued as $\gamma$ approaches zero in Fig.~\ref{Fig7}; instead, both
$\mbox{Im} \lec \alpha_{\calB3} \ric/\ko$ and $\mbox{Im} \lec \alpha_{\calB4}\ric /\ko$
are approximately equal to   $-0.01$ as $\gamma$ approaches zero. In the limiting case $\gamma \to 0$, material $\calB$ becomes a homogeneous material and the corresponding surface-wave solution represents a Dyakonov surface wave  \c{ESW_book,Dyakonov88}. Indeed, 
analytic formulas yield 
the angular existence domain $66.42 ^ \circ < \psi < 67.45^ \circ$, and the corresponding $q/\ko$ range 
$1.8850 < q/\ko < 1.8895$, for the corresponding Dyakonov wave that exists at $\gamma = 0$; the values of $\psi$ and $q/\ko$ for the surface-wave solution represented in Fig.~\ref{Fig7} lie within these ranges as $\gamma$ approaches zero.

\section{Closing remarks}
 
 The  theory underpinning the propagation of
Dyakonov--Tamm (DT) surface waves and Dyakonov--Tamm--Voigt (DTV) surface waves
was formulated for the canonical boundary-value problem involving the planar interface of a homogeneous  uniaxial dielectric material and a periodically nonhomogeneous isotropic dielectric material. Numerical studies were carried out with values
 corresponding to a realistic rugate filter \c{Bovard,Bau}
  for the highest and lowest refractive indexes of the periodically nonhomogeneous partnering material. Multiple DT surface waves were found to exist at a fixed propagation direction in the interface plane, depending upon the constitutive parameters of the partnering materials. These multiple solutions can be organized as
  continuous  branches  when regarded as functions of the propagation angle in the interface plane. Provided that the optic axis of the uniaxial partnering material lies in the interface plane, a single DTV surface-wave solution exists at  a unique propagation direction on each solution branch.
  
  The existence of multiple DT branch solutions~---~which is consistent with theoretical 
  \c{LP2007,DT_Reusch_pile,LF2,DT_temperature,High_phase_speed_DT},
  and experimental \c{PML2013,PMLHL2014} studies of DT surface waves supported by the planar interface of 
  a homogeneous isotropic  material and a periodically-nonhomogeneous anisotropic  material~---~is a feature that could be usefully exploited
  in optical sensing applications \c{LFjnp}, for example.
  
  The  unusual localization characteristics of DTV surface waves mirror those of
  Dyakonov--Voigt surface waves \c{MZL_PRSA,ZML_JOSAB} and surface-plasmon-polariton--Voigt waves \c{ZML_PRA}.
  While the existence of DTV surface waves is established theoretically herein 
  for an idealized scenario, i.e., the canonical boundary-value problem, further studies are required to elucidate the excitation of such waves and their propagation for partnering materials of finite thicknesses.

\vspace{5mm}

\noindent {\bf Acknowledgments.}
This work was supported  by
EPSRC (grant number EP/S00033X/1) and US NSF (grant number DMS-1619901).
AL thanks the Charles Godfrey Binder Endowment at the Pennsylvania State University  and  the Otto M{\o}nsted Foundation for partial support of his research endeavors.

\newpage

  \newpage

\newpage

\newpage

\newpage

\newpage

\newpage

\end{document}